\documentclass[aps,prd,nofootinbib,showkeys,floatfix,twocolumn,bibnotes,superscriptaddress,amsmath,amssymb,preprintnumbers]{revtex4-1}
\usepackage{booktabs,color,graphicx,verbatim}
\usepackage[colorlinks=true,citecolor=blue,filecolor=blue,linkcolor=blue,urlcolor=blue,pdftex]{hyperref}
\newcommand{\1}[1]{\, \mathrm{#1}} % unit(y ;-)
\newcommand{\mDM}{m_{\mathrm{DM}}}

\begin{document}

\title{Directly detecting sub-GeV dark matter with electrons from nuclear scattering}

\preprint{KCL-PH-TH/2017-54, TTK-17-43}

\author{Matthew J.\ Dolan}
\email{dolan@unimelb.edu.au}
\affiliation{ARC Centre of Excellence for Particle Physics at the Terascale, School of Physics, University of Melbourne, 3010, Australia}
\author{Felix Kahlhoefer}
\email{kahlhoefer@physik.rwth-aachen.de}
\affiliation{Institute for Theoretical Particle Physics and Cosmology (TTK), RWTH Aachen University, D-52056 Aachen, Germany}
\author{Christopher McCabe}
\email{christopher.mccabe@kcl.ac.uk}
\affiliation{Department of Physics, King's College London, Strand, London, WC2R 2LS, United Kingdom}

\begin{abstract}

  Dark matter (DM) particles with mass in the sub-GeV range are an attractive alternative to heavier weakly-interacting massive particles, but direct detection of such light particles is challenging. If however DM-nucleus scattering leads to ionisation of the recoiling atom, the resulting electron may be detected even if the nuclear recoil is unobservable. We demonstrate that including this effect significantly enhances direct detection sensitivity to sub-GeV DM. Existing experiments set world-leading limits, and future experiments may probe the cross sections relevant for thermal freeze-out.
  
\end{abstract}


\maketitle

\paragraph*{Introduction.---}%
Despite spectacular improvements in sensitivity over recent years, dark matter (DM) direct detection experiments have so far failed to observe conclusive evidence of a DM signal. While this may be interpreted as a failure of the paradigm of weakly-interacting massive particles (WIMPs)~\cite{Duerr:2016tmh,Escudero:2016gzx,Arcadi:2017kky}, an alternative explanation is that WIMPs are lighter than usually assumed and that the energy they can deposit in a detector is below current experimental thresholds (see~\cite{Battaglieri:2017aum} for a recent review). This consideration has led to increasing interest in experiments with lower thresholds for nuclear recoils, such as CRESST~\cite{Angloher:2015ewa, Angloher:2015eza,Angloher:2017sxg,Petricca:2017zdp}, DAMIC~\cite{Aguilar-Arevalo:2016ndq}, EDELWEISS~\cite{Arnaud:2017usi}, NEWS-G~\cite{Arnaud:2017bjh} or SuperCDMS~\cite{Agnese:2016cpb}, in experiments sensitive for electron recoils~\cite{Essig:2011nj,Graham:2012su,Essig:2015cda,Essig:2017kqs}, or indeed in the development of completely novel types of detectors~\cite{Guo:2013dt,Hochberg:2015pha,Hochberg:2015fth,Hochberg:2016ajh,Schutz:2016tid,Carter:2016wid,Hochberg:2016ntt,Derenzo:2016fse,Hochberg:2016sqx,Essig:2016crl,Knapen:2016cue,Bunting:2017net,Budnik:2017sbu,Tiffenberg:2017aac,Maris:2017xvi,Hochberg:2017wce,Fichet:2017bng}.

Direct detection experiments based on liquid xenon, on the other hand, are usually believed to be insensitive to nuclear recoils with sub-keV energy, corresponding to sub-GeV DM particles. This conclusion is  based on the implicit assumption that the electron cloud of the recoiling atom instantly follows the nucleus, so that ionisations and excitations are only produced subsequently, as the recoiling atom collides with surrounding xenon atoms. The resulting primary scintillation (S1) signal is then too small to be observable.

\begin{figure}
\centering
\includegraphics[width=0.99\columnwidth]{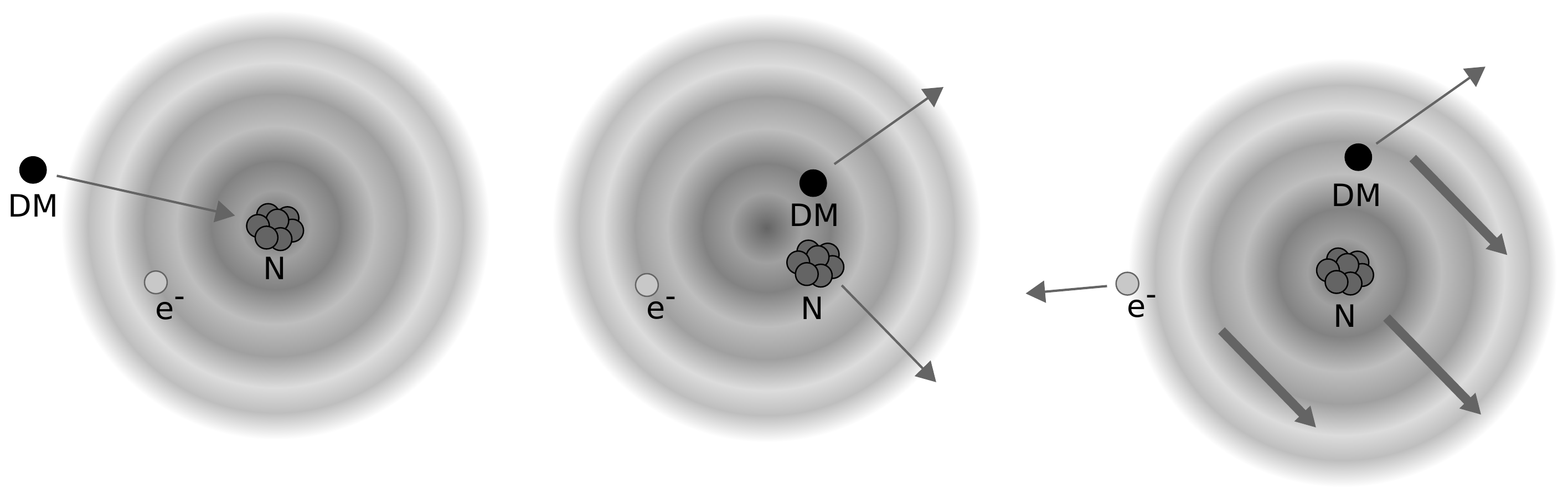}
\caption{\label{fig:figure1} Illustration of electron emission from nuclear recoils. If a DM particle scatters off a nucleus (panel 1), we can assume that immediately after the collision the nucleus moves relative to the surrounding electron cloud (panel 2). The electrons eventually catch up with the nucleus, but individual electrons may be left behind and are emitted, leading to ionisation of the recoiling atom (panel 3).}
\end{figure}

From neutron-nucleus scattering experiments it is however known that the sudden acceleration of a nucleus after a collision leads to excitations and ionisation of atomic electrons (see e.g.~\cite{Ruijgrok,Vegh,Baur,Pindzola,Sharma:2017fmo}). This effect, illustrated in figure~\ref{fig:figure1}, can lead to energetic $\gamma$-rays and ionisation electrons being produced from the primary interaction. The S1 signal is then much larger and the sensitivity of liquid xenon detectors is significantly enhanced. The case of $\gamma$-ray emission was investigated in detail in~\cite{Kouvaris:2016afs}. Ref.~\cite{McCabe:2017rln} furthermore showed that liquid xenon detectors such as LUX~\cite{Akerib:2016vxi}, XENON1T~\cite{Aprile:2017iyp} or PandaX~\cite{Cui:2017nnn} can distinguish such events from electron recoils, which constitute the main background.

Recently, ref.~\cite{Ibe:2017yqa} pointed out that the probability to ionise a recoiling atom is in fact substantially larger than the probability to obtain a $\gamma$-ray.
%, since in contrast to photons there is no momentum suppression for the emission of low-energy electrons.
This effect has been named ``Migdal effect'' in the DM literature~\cite{Vergados:2004bm,Moustakidis:2005gx,Bernabei:2007jz}, as the calculation makes use of the Migdal approximation~\cite{Landau:QM} that the electron cloud of the atom does not change during the nuclear recoil induced by the DM interaction (see figure~\ref{fig:figure1}). In the frame of the nucleus, this results in a simultaneous boost for all electrons, which can lead to excitation or ionisation of electrons.

In this letter, we further explore the formalism developed in~\cite{Ibe:2017yqa} and apply it to the case of liquid xenon detectors. We find that the sensitivity of this type of experiment in the sub-GeV mass range is significantly enhanced. By reinterpreting existing data from the LUX and XENON1T experiments, we obtain the strongest limit on DM with mass between $\sim 0.1-0.5$~GeV and comparable limits to CRESST-III~\cite{Petricca:2017zdp} between $\sim0.5-1$~GeV. A second central observation of our letter is that in scenarios where the DM couples with equal strength to electrons and protons (as in the case of interactions mediated by a dark photon with kinetic mixing~\cite{Foot:2004pa,Feng:2009mn}), experiments may be more sensitive to ionisation electrons resulting from nuclear recoils than from electron recoils. The search strategy considered in this paper can therefore probe unexplored parameter regions of dark photon models and future experiments may be sensitive to parameter space compatible with thermal freeze-out.

We begin by reviewing the central aspects of the Migdal effect and the relevant formulae from the literature. Next, we discuss the resulting signatures in liquid xenon detectors, considering the LUX experiment for concreteness. Finally, we present the resulting bounds on the parameter space of two interesting models of light DM.

%%%%%%%%%%%%%%%%%%%%%%%%%%%%%%%%%%%%%%%%%%%%%%%%%%%%%%%
\smallskip
\paragraph*{Ionization electrons from nuclear recoils.---}%
In this section we summarize the key results from~\cite{Ibe:2017yqa} needed for calculating the signatures of light DM scattering on nuclei in direct detection experiments. The differential event rate for DM-nucleus scattering $R_\text{nr}$ with respect to the nuclear recoil energy $E_\mathrm{R}$ and the DM velocity $v$ is given by
\begin{equation}
 \frac{\mathrm{d}^2 R_\text{nr}}{\mathrm{d}E_\mathrm{R}\,\mathrm{d}v} = \frac{\rho \, \sigma_N}{2 \, \mu_N^2 \, \mDM} \frac{f(v)}{v} 
\; ,
\end{equation}
where $\rho$ denotes the local DM density, $\sigma_N$ the DM-nucleus scattering cross section\footnote{We have absorbed the coherent enhancement factor into our definition of $\sigma_N$.}, $\mDM$ the DM mass, $\mu_N = m_N \, \mDM / (m_N + \mDM)$ the DM-nucleus reduced mass and $f(v) = \int v^2 \, f(\mathbf{v}) \, \mathrm{d}\Omega_v$ the DM speed distribution in the laboratory frame~\cite{McCabe:2013kea}. We neglect nuclear form factors since we are only interested in small momentum transfers. The differential event rate for a nuclear recoil of energy $E_\mathrm{R}$ to be accompanied by an ionisation electron with energy $E_e$ is
\begin{equation}
 \frac{\mathrm{d}^3 R_\text{ion}}{\mathrm{d}E_\mathrm{R}\,\mathrm{d}E_e\,\mathrm{d}v} = \frac{\mathrm{d}^2R_\text{nr}}{\mathrm{d}E_\mathrm{R}\,\mathrm{d}v} \times \left|Z_\text{ion}(E_\mathrm{R},E_e)\right|^2 \; ,
 \label{eq:fulldR}
\end{equation}
where the transition rate is given by
\begin{equation}
\left|Z_\text{ion} (E_R, E_e) \right|^2 = \sum_{nl} \frac{1}{2 \pi} \frac{\mathrm{d} p^c_{q_e}(nl \to E_e)}{\mathrm{d}E_e}\; .
\end{equation}
In this expression $n$ and $l$ denote the initial quantum numbers of the electron being emitted, $q_e = m_e \sqrt{2 \, E_\mathrm{R} / m_N}$ is the momentum of each electron in the rest frame of the nucleus immediately after the scattering process, and $p^c_{q_e}(nl \to E_e)$ quantifies the probability to emit an electron with final kinetic energy $E_e$. We can make the dependence of $p^c_{q_e}(nl \to E_e)$ on $q_e$ explicit by writing
\begin{equation}
p^c_{q_e}(nl \to E_e) = \left(\frac{q_e}{v_\text{ref} \, m_e}\right) p^c_{v_\text{ref}}(nl \to E_e) \; ,
\end{equation}
where $v_\text{ref}$ is a fixed reference velocity. The functions $p^c_{v_\text{ref}}(nl \to E_e)$ depend on the target material under consideration. We use the functions from ref.~\cite{Ibe:2017yqa}, which have been calculated taking $v_\text{ref} = 10^{-3}$.

If the emitted electron comes from an inner orbital, the remaining ion will be in an excited state. To return to the ground state, further electronic energy will be released in the form of photons or additional electrons.\footnote{In contrast, the probability to obtain double ionisation from the Migdal effect itself is exceedingly small~\cite{Feist,Liertzer}.} The total electronic energy deposited in the detector is hence approximately given by $E_\text{EM} = E_e + E_{nl}$, where $E_{nl}$ is the (positive) binding energy of the electron before emission.
%and we assume that the binding energy of the electrons in the outermost orbitals are negligible.
%The nuclear recoil can also lead to the atom being in double or higher states of ionisation. Existing results for the helium atom show that double ionisation is $\mathcal{O}(10^{-3})$ less likely than single ionisation~\cite{Feist,Liertzer}. Accordingly, we neglect this possibility in this work.

We integrate eq.~\eqref{eq:fulldR} over the nuclear recoil energy and the DM velocity to calculate the energy spectrum, including only those combinations of $E_\mathrm{R}$, $E_\text{EM}$ and $v$ that satisfy energy and momentum conservation. The resulting calculation is identical to the case of inelastic DM~\cite{TuckerSmith:2001hy}, with the DM mass splitting $\Delta m$ being replaced by the total electronic energy $E_\text{EM}$.\footnote{We neglect the difference in mass between the original atom and the recoiling excited state.} We find
\begin{equation}
v_{\rm{min}} = \sqrt{\frac{m_N E_{\rm{R}}}{2 \mu^2}} + \frac{E_\text{EM}}{\sqrt{2 m_N E_{\rm{R}}}} \; .
\end{equation}

The maximum electronic and nuclear recoil energy for a given DM mass are given by 
\begin{equation}
 E_\text{R,max} = \frac{2\,\mu_N^2\,v_\text{max}^2}{m_N} \; , \quad E_\text{EM,max} = \frac{\mu_N \, v_\text{max}^2}{2} \; .
\end{equation}
For $v_\text{max} \approx 800 \, \mathrm{km/s}$, $\mDM \ll m_N$ (and hence $\mu_N \approx \mDM$), we generically find $E_\text{EM,max} \gg  E_\text{R,max}$. For concreteness, for $\mDM = 0.5\,\mathrm{GeV}$ and $m_N = 120\,\mathrm{GeV}$ (the approximate xenon atom mass), we find $E_\text{R,max} \approx 0.03\,\mathrm{keV}$ while $E_\text{EM,max} \approx 1.8\;\text{keV}$. The electronic energy is therefore much easier to detect than the nuclear recoil energy.

%%%%%%%%%%%%%%%%%%%%%%%%%%%%%%%%%%%%%%%%%%%%%%%%%%%%%%%
\smallskip
\paragraph*{Sensitivity of liquid xenon detectors.---}%
Having obtained the relevant formulae for the distribution of electronic and nuclear recoil energy at the interaction point where the DM-nucleus scattering occurs, we now convert these energies into observables accessible for direct detection experiments. The focus of this discussion will be on liquid xenon detectors, but we note that the dominance of the electronic energy~$E_\text{EM}$ resulting from the Migdal effect is not limited to xenon. These detectors convert the atomic excitations and ionisations at the interaction point into a primary~(S1) and a secondary~(S2) scintillation signal~\cite{Chepel:2012sj}. A specific detector can be characterized by two functions: $\text{pdf}(\text{S1,S2}|E_\mathrm{R},E_\mathrm{EM})$ quantifies the probability to obtain specific~S1 and~S2 values for given $E_\mathrm{R}$ and $E_\mathrm{EM}$; and $\epsilon(\text{S1,S2})$ quantifies the probability that a signal with given~S1 and~S2 will be detected and will satisfy all selection cuts. Using these two functions, we can write
\begin{align}
 \frac{\mathrm{d}^2 R}{\mathrm{d}\text{S1} \, \mathrm{d}\text{S2}} = \epsilon(\text{S1,S2}) \int & \mathrm{d}E_\mathrm{R} \, \mathrm{d}E_\text{EM}  \frac{\mathrm{d}^2 R}{\mathrm{d}E_\mathrm{R} \, \mathrm{d}E_\text{EM}} \nonumber \\ & \times \text{pdf}(\text{S1,S2}|E_\mathrm{R},E_\mathrm{EM}) \; ,
 \label{eq:d2RS1S2}
\end{align}
where we have now expressed the differential event rate from eq.~\eqref{eq:fulldR} in terms of $E_{\rm{EM}}$.

For sub-GeV DM particles, nuclear recoil energies are below $\mathcal{O}(0.1)$~keV. The scintillation and ionisation yield for such small energies have not yet been measured in liquid xenon, but theoretical arguments predict the resulting signals to be very small~\cite{Sorensen:2014sla}. We conservatively neglect this contribution and assume that only electronic energy contributes to the S1 and S2 signals, such that $\text{pdf}(\text{S1,S2}|E_\mathrm{R},E_\mathrm{EM}) = \text{pdf}(\text{S1,S2}|E_\mathrm{EM})$ and the integration over $E_\mathrm{R}$ in eq.~(\ref{eq:d2RS1S2}) can be immediately performed.

Refs.~\cite{McCabe:2015eia, McCabe:2017rln} and Appendix~\ref{app:1} discuss how we determine $\text{pdf}(\text{S1,S2}|E_\mathrm{EM})$ and $\epsilon(\text{S1,S2})$ for liquid xenon experiments using a Monte Carlo simulation of the detector based on the Noble Element Simulation Technique (NEST)~\cite{Szydagis:2011tk,Szydagis:2013sih,Lenardo:2014cva}. For given $E_\text{EM}$, the mean~S1 and~S2 signals can be written as $\text{S1} = g_1 \, L_y \, E_\text{EM}$ and $\text{S2} = g_2 \, Q_y \, E_\text{EM}$, respectively, where $g_{1,2}$ are detector-dependent gain factors and $L_y$ and $Q_y$ are properties of liquid xenon determined from calibration data~\cite{Akerib:2015wdi,Goetzke:2016lfg,Boulton:2017hub,Akerib:2017hph,Aprile:2017xxh}. The Monte Carlo simulation then determines the probability for fluctuations in the number of excited and ionised atoms produced initially, recombination fluctuations as well as finite extraction and detection efficiencies.

As pointed out in~\cite{McCabe:2017rln}, these fluctuations play a crucial role for the sensitivity of liquid xenon detectors to sub-GeV DM particles, because the mean S1 signal expected from the scattering of light DM particles lies below the detection threshold of typical detectors. Thus, the signal can only be observed in the case of an upward fluctuation of the S1 signal. Moreover, events with unusually large S1 signal have the advantage that they do not look like typical electronic recoils.
%, which are the dominant source of background in these detectors.
Instead, they look more similar to nuclear recoils, which have a smaller ratio of S2/S1 than electron recoils. This feature makes it possible to distinguish between the expected signal and the main sources of backgrounds.

We focus on the LUX~\cite{Akerib:2015rjg,Akerib:2016vxi} and XENON1T~\cite{Aprile:2017iyp,Aprile:2018dbl} experiments, and following~\cite{McCabe:2017rln}, implement two different approaches to estimate the sensitivity to sub-GeV DM, adopting the astrophysical parameters used in~\cite{McCabe:2017rln} for our analysis.
%\footnote{Although XENON1T and PandaX have larger exposures than LUX, the analysis cuts and thresholds mean that LUX is marginally more sensitive at low energies.} 
For the cut-and-count (CC) approach we determine the region in S1-S2 space that contains 90\% of the DM events passing all cuts and count the number of observed events in this region. A limit can then be set by assuming all these events to be signal events and calculating a Poisson upper bound on the expected number of events (at 90\% confidence level). A more powerful approach is the profile likelihood method (PLR), which takes into account the likelihood for signal and background at each observed event~\cite{Barlow:1990vc,Cowan:2010js}. In contrast to the CC approach, the PLR method requires a background model. For LUX, our model consists of two components: a component flat in energy from electronic recoils and a component from the decays of $^{127}$Xe. Further details are given in~\cite{McCabe:2017rln}. For XENON1T, modelling the surface background is beyond the scope of this work so we only show results for a CC approach. With this approach we obtain a stronger bound from the first run (SR0) than from the second run (SR1) due to the smaller number of background events. We expect that a PLR analysis of SR1 would further improve the XENON1T exclusion limit.

We also estimate the expected sensitivity of the LZ experiment~\cite{Mount:2017qzi}, and note that a similar sensitivity should be expected with XENONnT~\cite{Aprile:2015uzo}. We show results assuming 1000~days of data taking and a 5.6~tonne target volume using the detector parameters sets discussed in~\cite{Akerib:2018lyp}. 
%The `goal' parameter set results in a higher sensitivity primarily owing to a higher value of $g_1$, a lower trigger threshold and a lower radon background rate.

%%%%%%%%%%%%%%%%%%%%%%%%%%%%%%%%%%%%%%%%%%%%%%%%%%%%%%%
\smallskip
\paragraph*{Results.---}%

\begin{figure}[t]
\centering
\includegraphics[width=\columnwidth]{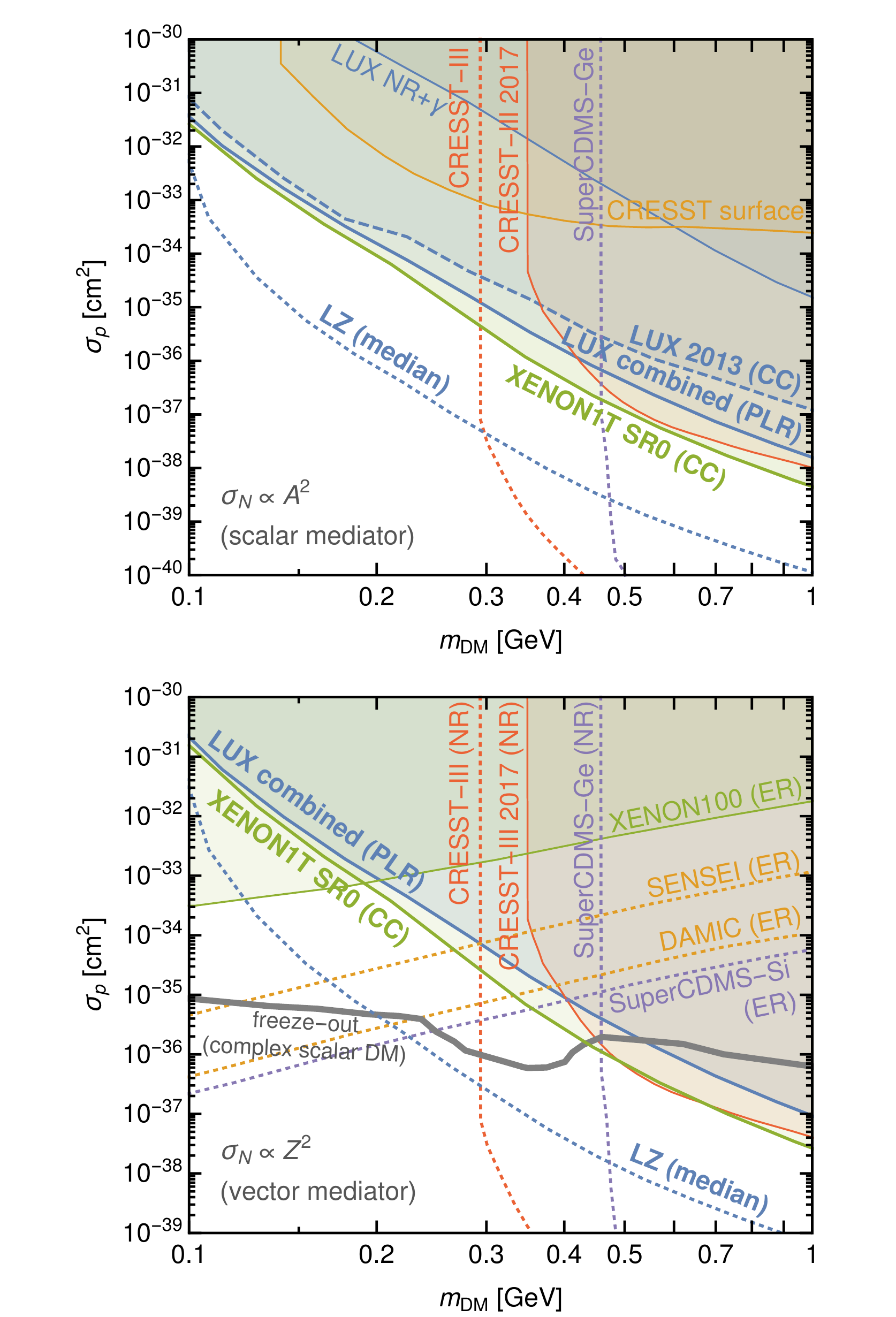}
\caption{\label{fig:results}Exclusion limits (solid lines) and projected sensitivities (dotted lines) for sub-GeV DM. The bounds resulting from our analysis of electron emission after nuclear recoils are shown in blue for LUX (using two different statistical methods) and in green for XENON1T. The upper panel considers the case of a scalar mediator (couplings proportional to mass), the lower panel considers the case of a vector mediator (couplings proportional to charge). In the latter case, there are constraints both from experiments looking for nuclear recoils (NR) and from experiments looking for electron recoils (ER). The bounds from the CRESST surface run, LUX NR+$\gamma$ and LUX 2013 (CC) have been omitted in the lower panel for clarity. We also show the parameter combinations that yield the observed DM relic abundance for one specific model (complex scalar DM). See text for details.}
\end{figure}

We present the results of our analysis for DM particles interacting with nuclei via two different types of mediators: scalars and vectors. The difference between these two cases is how the mediator couples to Standard Model (SM) particles~\cite{Kaplinghat:2013yxa}. For a scalar mediator these couplings generically arise from mixing with the SM Higgs boson, so that the mediator couples to SM particles proportional to their mass. In particular, the DM-nucleus cross section for nuclei with mass number~$A$ is enhanced by a coherence factor, $\sigma_N = A^2 \, \sigma_p \, \mu_N^2/\mu_p^2$, where $\sigma_p$ and $\mu_p$ are the DM-proton cross section and reduced mass, respectively. Furthermore, couplings to electrons are negligible. Models with sub-GeV DM particles and (light) scalar mediators have for example been considered recently in the context of self-interacting DM~\cite{Kahlhoefer:2017umn,Kahlhoefer:2017ddj}.

For vector mediators, couplings to SM particles arise from kinetic mixing with the photon. As a result, couplings to SM particles are expected to be proportional to their electromagnetic charge, leading to a $Z^2$ enhancement for scattering on nuclei with charge number $Z$, i.e.\ $\sigma_N = Z^2 \, \sigma_p \, \mu_N^2/\mu_p^2$, and comparable couplings to protons and electrons. These so-called dark photon models have been studied extensively in the literature~\cite{An:2014twa,Essig:2015cda,Essig:2017kqs,Essig:2013vha,Lees:2017lec,Andreas:2012mt,Izaguirre:2013uxa,Batell:2014mga}.
%and there has been much work on the possibility to search for them in direct detection experiments~\cite{An:2014twa,Essig:2015cda,Essig:2017kqs}, at low-energy colliders~\cite{Essig:2013vha,Lees:2017lec} and at beam-dump experiments~\cite{Andreas:2012mt,Izaguirre:2013uxa,Batell:2014mga}.

We emphasize that even if the couplings of DM to protons and electrons are comparable, the corresponding scattering cross sections are not, because they are proportional to the reduced mass squared~\cite{Essig:2011nj}. For sub-GeV DM particles, one finds $\mu_p \approx \mDM$ and $\mu_e \approx m_e$, so scattering on nucleons is enhanced by a factor $\mDM^2 / m_e^2$. For heavy atoms, the coherence factor for the nucleus leads to an additional enhancement in spite of the larger number density of electrons. Thus, the probability for DM particles in the mass range 0.1--1 GeV to scatter on nuclei is many orders of magnitude larger than the probability to scatter on electrons. Most of these scattering processes will be unobserved, but even a small fraction of events with ionisation electrons are sufficient to obtain strong constraints.

Our results are summarized in figure~\ref{fig:results}, focusing on the mass range $0.1 \, \mathrm{GeV} \leq \mDM \leq 1\,\mathrm{GeV}$. The upper panel shows the case of a scalar mediator (Higgs mixing), the lower panel the case of a vector mediator (kinetic mixing). An additional assumption in both plots is that the mediator is sufficiently heavy that the scattering can be described by contact interactions, which is the case for $m_\text{med} \gtrsim 10\,\mathrm{MeV}$~\cite{Kahlhoefer:2017ddj}. No further assumptions are needed to compare the constraints from different direct detection experiments looking for nuclear or electron recoils.\footnote{We do not show constraints from hidden photon searches at BaBar~\cite{Lees:2017lec}, which require the assumption of a specific ratio between the DM mass and the mediator mass.} Projected sensitivities are taken from~\cite{Agnese:2016cpb} for SuperCDMS-Ge, from~\cite{Kahlhoefer:2017ddj} for CRESST-III and from~\cite{Battaglieri:2017aum} for SENSEI, DAMIC and SuperCDMS-Si.

We find that in both of the cases considered, the sensitivity of current liquid xenon detectors for sub-GeV DM can compete with other existing and proposed strategies. In the case of a scalar mediator, only experiments sensitive to nuclear recoils give relevant constraints. We find that XENON1T gives the world-leading exclusion limit across the entire mass range under consideration. In particular, we find XENON1T to be more constraining than the first analysis of CRESST-III~\cite{Petricca:2017zdp} even for $\mDM \gtrsim 0.5 \, \mathrm{GeV}$. For lower DM masses, we significantly improve the LUX bound obtained from $\gamma$-rays emitted in nuclear scattering processes~\cite{McCabe:2017rln}, and from the CRESST~2017 surface run~\cite{Angloher:2017sxg}.

In the case of vector mediators, constraints from LUX and XENON1T are slightly weakened relative to the ones from CRESST due to a smaller difference in the enhancement factors. In addition, there are now strong constraints from searches for electron scattering in XENON100~\cite{Aprile:2016wwo,Essig:2017kqs}. Nevertheless, we observe that searches for electrons emitted for nuclear recoils in liquid xenon detectors set competitive bounds for DM masses around 200--400 MeV. LZ can significantly improve upon these bounds in the future and provide complementary constraints to alternative electron-recoil strategies proposed to search for sub-GeV DM~\cite{Battaglieri:2017aum}.

To put our results into context, it is instructive to compare the sensitivity of direct detection experiments to the parameter regions where the DM particle can be a thermal relic that obtains its abundance via the freeze-out mechanism. Such a comparison is necessarily model dependent, but the number of possibilities is limited by strong constraints on sub-GeV WIMPs~\cite{Boehm:2013jpa,Knapen:2017xzo}. Here we focus on one viable scenario, namely a vector mediator and complex scalar DM~\cite{Essig:2015cda}. If the mediator mass is sufficiently large, $m_\text{med} > 2 \mDM$, the DM relic abundance probes the same combination of parameters as direct detection experiments. In other words, for each value of $\mDM$ there is a unique value of~$\sigma_p$ corresponding to the observed relic abundance~\cite{Ade:2015xua}. The scattering cross sections obtained in this way are indicated by the grey band in the lower panel of figure~\ref{fig:results}. We observe that current bounds exclude the simplest realization of thermal freeze-out in the model that we consider for $\mDM > 450\,\mathrm{MeV}$, while the next generation of direct detection experiments is expected to probe the relevant cross sections across the entire mass range of interest.

%%%%%%%%%%%%%%%%%%%%%%%%%%%%%%%%%%%%%%%%%%%%%%%%%%%%%%%
\smallskip
\paragraph*{Conclusions.---}%
The sub-GeV mass range represents a new frontier in the search for particle DM. While direct detection experiments are often considered insensitive to nuclear recoils induced by scattering of light DM particles, we have shown that electrons emitted from recoiling atoms significantly boost the signal, leading to an enhanced sensitivity for, and new bounds on, the DM-nucleus scattering cross-section. In liquid xenon detectors the sensitivity to these signals is further enhanced by the possibility of upward fluctuations in the S1 signal, leading to the possibility to distinguish between signal and background.

In the present work we have focused on the LUX experiment as an example of existing technology and the LZ experiment as illustration of the power of next-generation experiments. Nevertheless, the same physics will be relevant for other direct detection experiments sensitive to electronic recoils.
Moreover, our results may also be applied to coherent neutrino-nucleus scattering~\cite{Akimov:2017ade,Ibe:2017yqa} and to the interpretation of calibration data based on neutron-nucleus scattering~\cite{Verbus:2016sgw,Barbeau:2007qh}.

In conclusion, we emphasize that direct detection experiments are only just beginning to probe the interesting parameter space for sub-GeV WIMPs. Significant improvements of sensitivity are required in order to probe the cross sections favoured by the freeze-out mechanism. To achieve this goal we need both a dedicated effort to build new types of direct detection experiments with very low thresholds, and further theoretical developments to better understand the ways in which DM particles can lead to observable signals in conventional direct detection experiments.

\bigskip
%\vfill 
\begin{acknowledgments}
We thank Josef Pradler and Tien-Tien Yu for discussions and Florian Reindl for correspondence. MJD is supported by the Australian Research Council and thanks King's College London for hospitality during the commencement of this project. FK is supported by the DFG Emmy Noether Grant No.\ KA 4662/1-1. CM is supported by the Science and Technology Facilities Council (STFC) Grant ST/N004663/1. 
\end{acknowledgments}

\bigskip

\appendix

\section{Modelling electronic recoil events in a xenon detector} 
\label{app:1}

A basic input for all of our detector simulations is the electron yield $Q_y$. This is related to the observable (mean) S2 signal through the relation $\mathrm{S2}=g_2\, Q_y \,E_{\mathrm{EM}}$, where $g_2$ is a detector-dependent gain factor and $E_{\mathrm{EM}}$ is the electronic recoil energy. For electronic recoils, $Q_y$ is also simply related to the photon yield $L_y$ through the relation
\begin{equation}
\frac{1}{W} = L_y + Q_y\;,
\end{equation}
where $W=13.7$ is the work function for xenon~\cite{Chepel:2012sj}. The photon yield is related to the observable (mean) S1 signal through the relation $\mathrm{S1}=g_1 \,L_y\, E_{\mathrm{EM}}$, where $g_1$ is another detector-dependent gain factor.

\begin{figure}[!t]
\centering
\includegraphics[width=0.95\columnwidth]{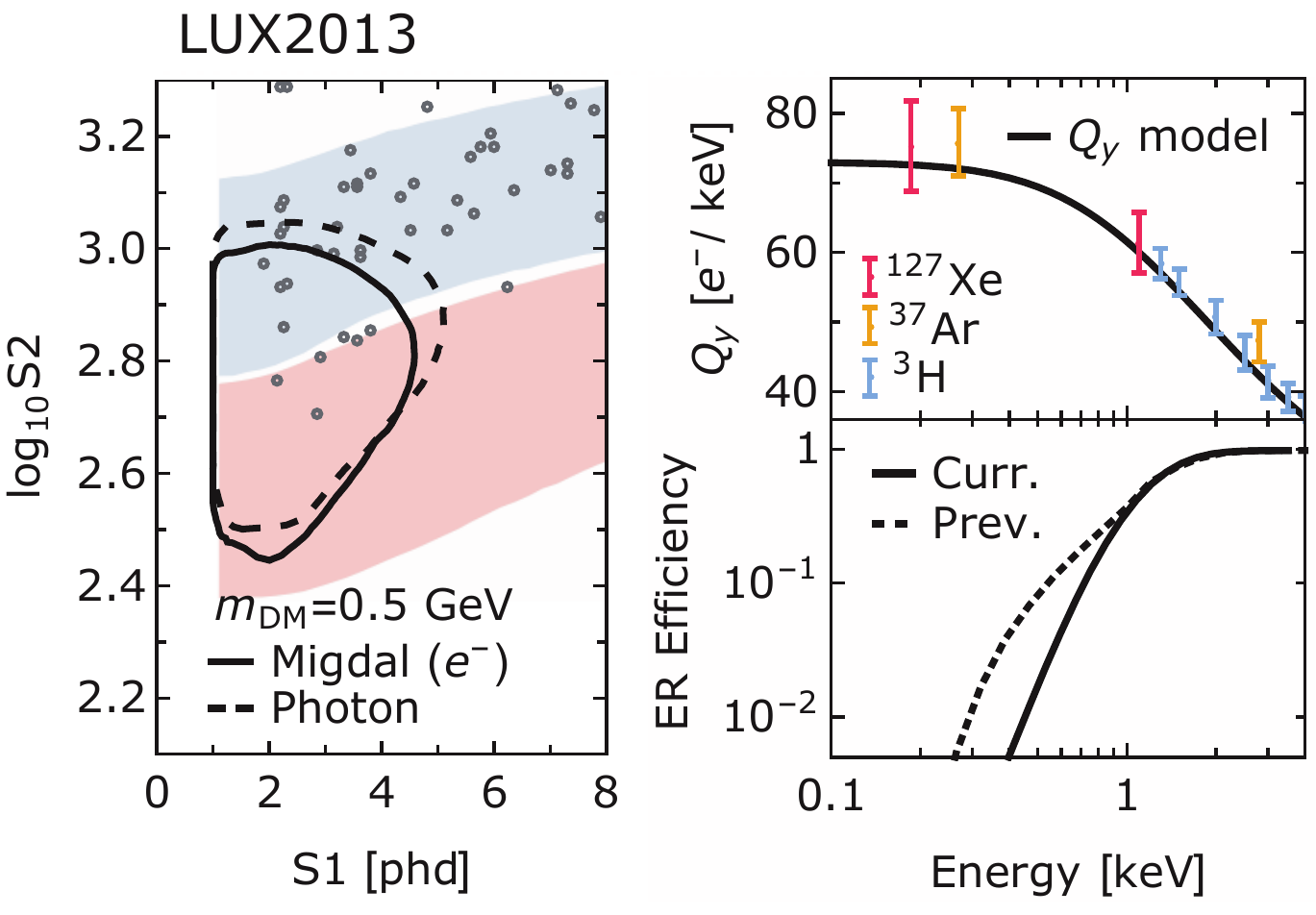}
\includegraphics[width=0.95\columnwidth]{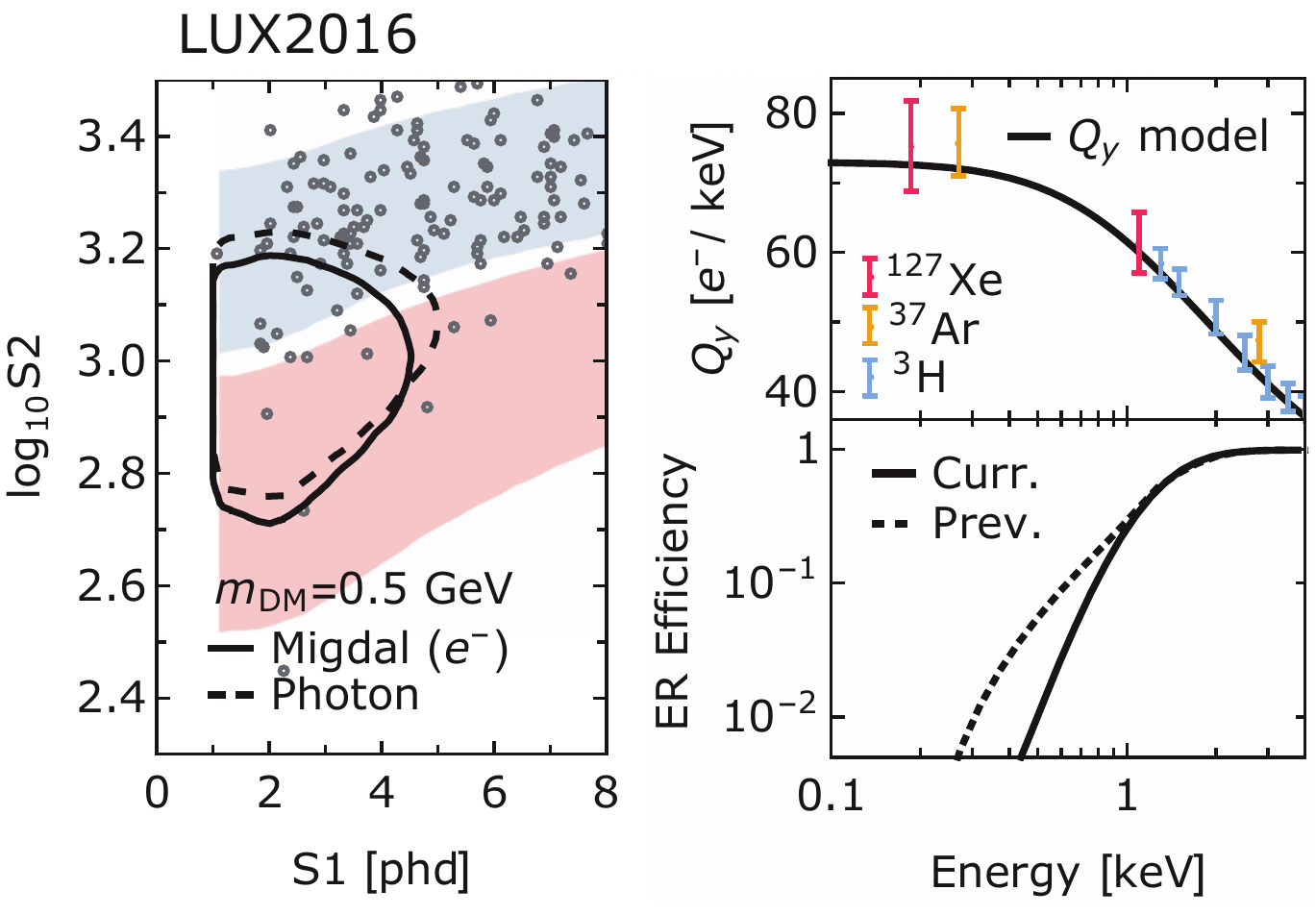}
\caption{Left panels: Blue and red shaded regions show the electronic and nuclear recoil bands respectively, obtained from our detector simulations. The black solid and black dashed show regions that contain $90\%$ of DM events when $m_{\mathrm{DM}}=0.5~\mathrm{GeV}$ for the Migdal effect and for $\text{nuclear recoils} + \gamma$ emission, respectively. The grey points show the events measured during the LUX2013 (upper panel) and LUX2016 (lower panel) periods of data taking, respectively. For LUX2016, we exclude events within 1~cm of the radial fiducial volume. Upper right panels: The black solid line shows the parameterisation of the electron yield $Q_y$ used in our detector simulations. The red, blue and orange data points show low-energy calibration data measured with the LUX and PIXeY experiments. Lower right panels: The detection efficiency for electronic recoils as a function of energy. The solid black line shows the efficiency obtained from our simulations in this work. The dashed line shows the efficiency used in a previous study (ref.~\cite{McCabe:2017rln}) of $\text{nuclear recoils} + \gamma$ emission.}
\label{fig:LUX}
\end{figure}

\begin{figure}[!t]
\centering
\includegraphics[width=0.95\columnwidth]{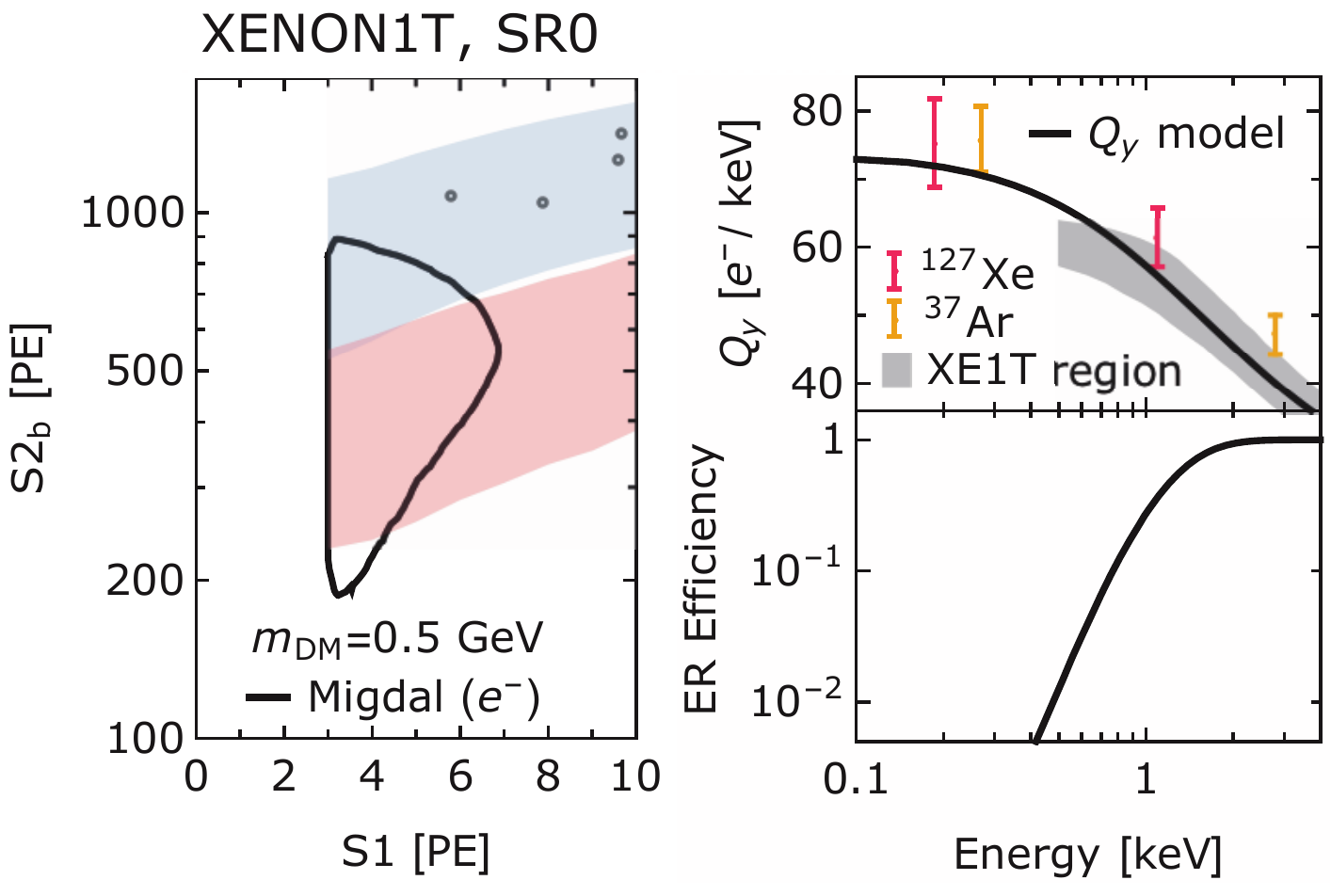}
\includegraphics[width=0.95\columnwidth]{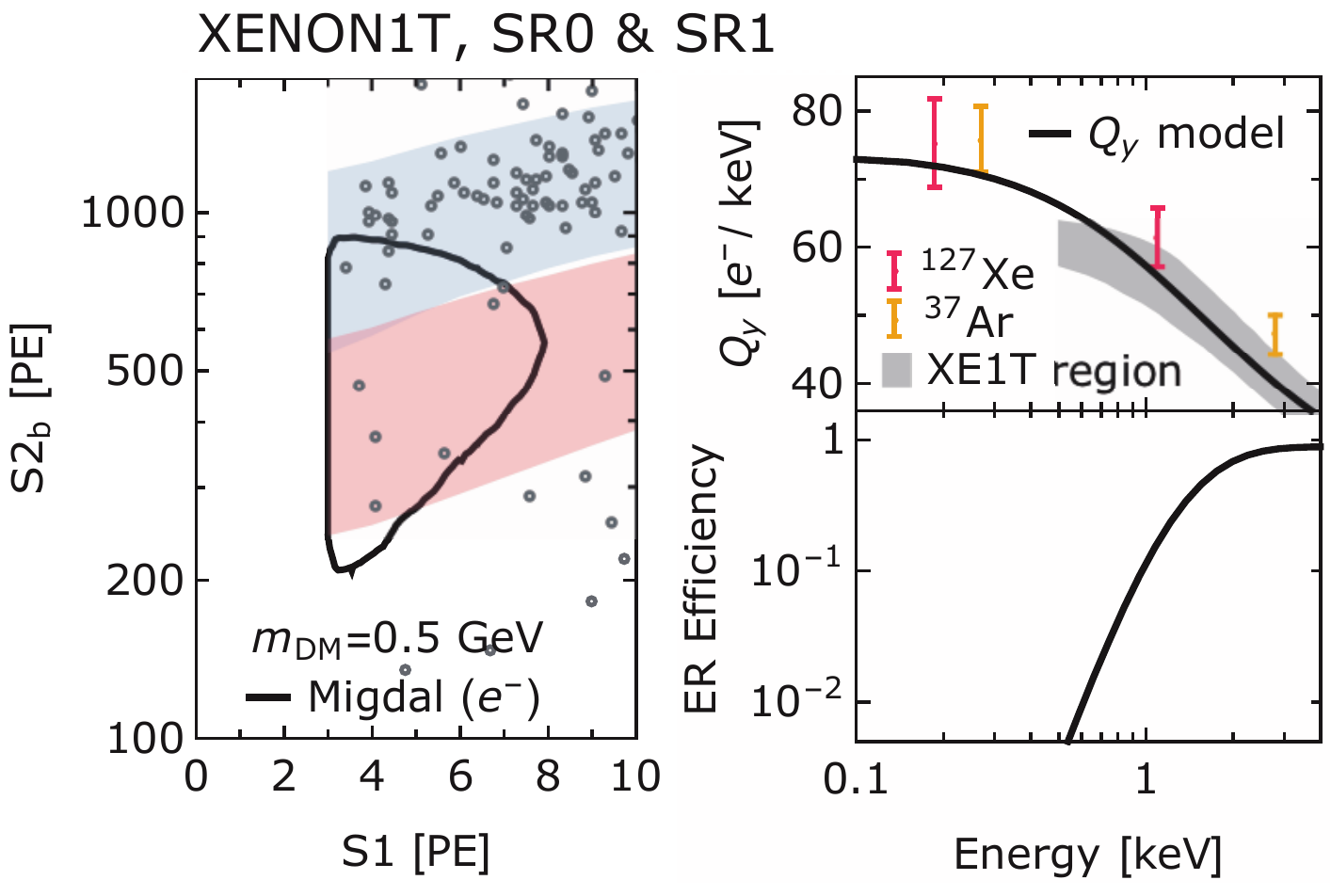}
\caption{Similar to figure~\ref{fig:LUX}. Left panel: The electronic and nuclear recoil bands, together with the Migdal signal region are plotted in the S1 -- S2$_{\mathrm{b}}$ plane. Very few events were observed during the first DM search (`SR0'). We include all events observed within the 1.3~tonne fiducial volume of the total exposure (`SR0'+`SR1').  Upper right panels: We compare our $Q_y$ parameterisations with the XENON collaboration's best estimate of $Q_y$ (grey shaded regions).\label{fig:XE1T}}
\end{figure}

\begin{figure}[!t]
\includegraphics[width=0.95\columnwidth]{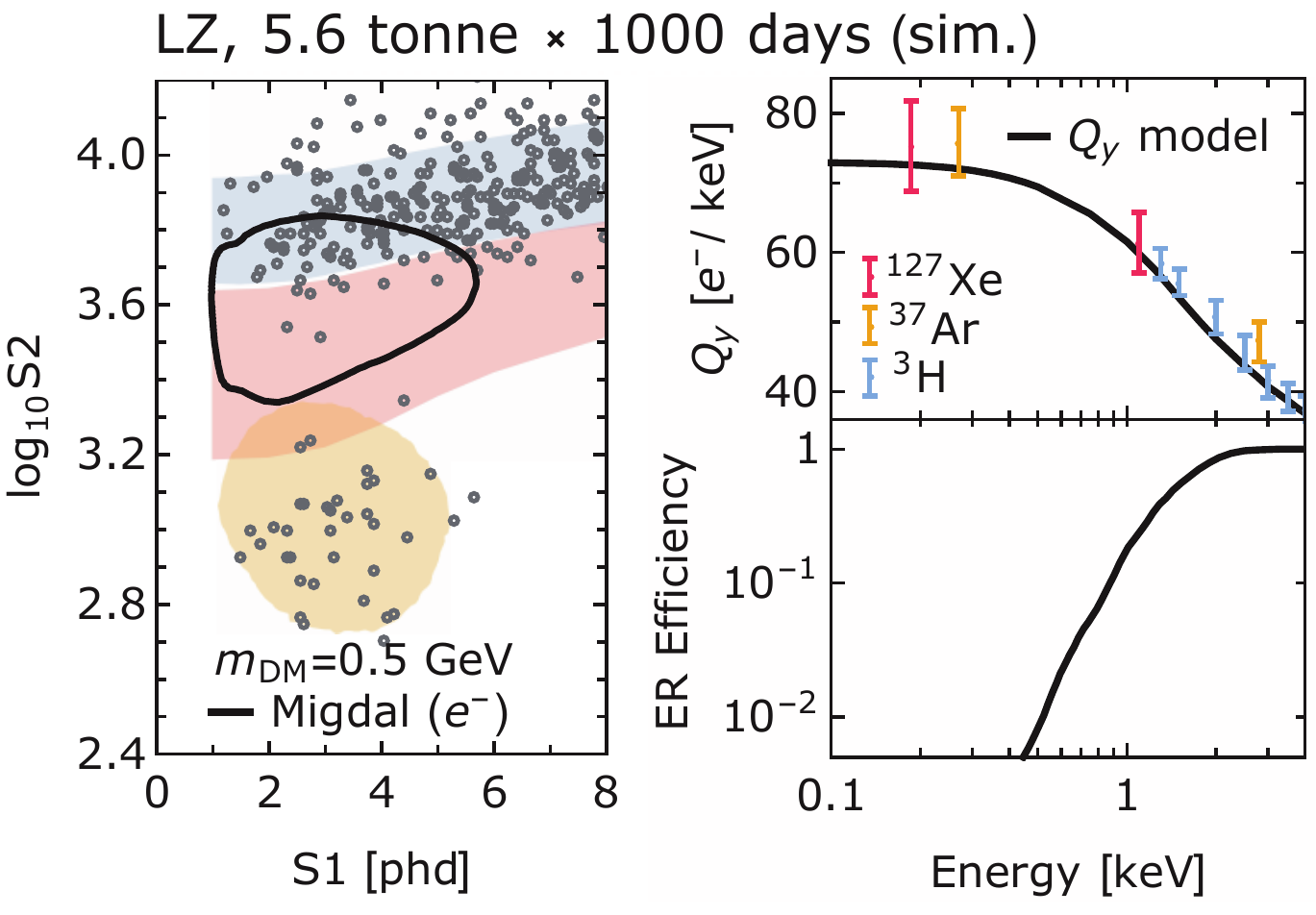}
\caption{Similar to figure~\ref{fig:LUX}. In the left panel of this figure, the orange region indicates the $90\%$ background region from $^8\mathrm{B}$ and hep solar neutrino induced nuclear recoils. The grey points show simulated data for a background-only $5.6~\mathrm{tonne}\times1000~\mathrm{day}$ exposure. \label{fig:LZ}}
\end{figure}

The upper right panels in figures~\ref{fig:LUX} and~\ref{fig:XE1T} show our parameterisations of $Q_y$ that we use for each of our detector simulations in the low-energy regime most relevant for our analysis. For LUX2013, LUX2016 and LZ (figures~\ref{fig:LUX} and~\ref{fig:LZ}), we fit to the low-energy $^{127}\mathrm{Xe}$ data~\cite{Akerib:2017hph} and tritium calibration data~\cite{Akerib:2015wdi} collected by LUX and to the PIXeY data from decays of $^{37}\mathrm{Ar}$~\cite{Boulton:2017hub}. Although LZ is projected to operate with a different drift field, the low-energy dependence of $Q_y$ on the drift field is small (for recent results, see e.g.~\cite{fields}). For XENON1T, we fit to the lowest energy $^{127}\mathrm{Xe}$ and $^{37}\mathrm{Ar}$ points but at higher energies, match our $Q_y$ values to the credible regions determined by the XENON1T collaboration~\cite{Aprile:2017xxh}. These credible regions are shaded in grey in the upper right panel of figure~\ref{fig:XE1T}.

The electron yield $Q_y$ is a basic input for calculating the mean S1 and S2 signals. To include fluctuations in our model, we follow the microphysics model introduced by NEST~\cite{Szydagis:2011tk}. A recent succinct summary of this model is provided in ref.~\cite{Aprile:2017xxh}. In addition to $Q_y$, we must specify the exciton-to-ion ratio $\alpha$, the mean recombination probability $r$ and the recombination fluctuation $\Delta r$. We use the value $\alpha=0.12$, consistent with the expected range of 0.06 -- 0.2~\cite{Szydagis:2011tk}, while our energy-dependent parameterisations for $r$ and $\Delta r$ lie within the credible regions given in ref.~\cite{Aprile:2017xxh}. Finally, we also allow for variations in the light collection as a function of depth in the detector and allow for electrons to be captured as they drift from the interaction point to the top of the detector. For LUX, we take specific detector parameters, such as PMT acceptance, relative light collection efficiency, single electron size and electron lifetime from ref.~\cite{Akerib:2017vbi}. For our LZ simulation, we take detector parameters and follow the procedure in ref.~\cite{Akerib:2018lyp} while for XENON1T, we make use of the information in refs.~\cite{Aprile:2017xxh, aablers}. Our detector simulations accurately reproduce the electronic and nuclear recoils bands for the respectively experiments. The results from our simulations are shown as the blue and red shaded regions in the left panels of figures~\ref{fig:LUX} and~\ref{fig:XE1T}, respectively.

In the bottom right panels of figs~\ref{fig:LUX} to~\ref{fig:XE1T}, we show the efficiency of detecting a signal as a function of energy for electronic recoil events. All of the resulting efficiencies are reasonably similar. This is a reflection of the similar light collection efficiencies in the different experiments, which is what dominates the efficiency at low energies. Of particular note is the new efficiency that we use for LUX2013 and LUX2016. The solid and dashed lines in the lower right panel of figure~\ref{fig:LUX} compare the efficiency from the improved detector simulation used in this analysis with the result from a previous study~\cite{McCabe:2017rln}. We have used this improved efficiency to recalculate the LUX~$\mathrm{NR} + \gamma$ limit~\cite{McCabe:2017rln}.

Finally, the black solid contour in the left panels of figs~\ref{fig:LUX} to~\ref{fig:XE1T} shows the region that contains 90\% of the signal events in the S1 -- S2 plane for $\mDM=0.5~\mathrm{GeV}$. These signal regions lie preferentially below the usual electronic recoil band. In the left panel of figure~\ref{fig:LUX}, we also show the signal region for the LUX~$\mathrm{NR} + \gamma$ (black dashed line) for $\mDM=0.5~\mathrm{GeV}$. This is similar to the signal region from the Migdal effect. The small differences arise from the slightly different shapes of the recoil spectra. The orange region in the lower left panel of figure~\ref{fig:LZ} shows the signal region expected from $^8\mathrm{B}$ and hep solar neutrino induced nuclear recoils.


\newpage

\begin{thebibliography}{91}%
\makeatletter
\providecommand \@ifxundefined [1]{%
 \@ifx{#1\undefined}
}%
\providecommand \@ifnum [1]{%
 \ifnum #1\expandafter \@firstoftwo
 \else \expandafter \@secondoftwo
 \fi
}%
\providecommand \@ifx [1]{%
 \ifx #1\expandafter \@firstoftwo
 \else \expandafter \@secondoftwo
 \fi
}%
\providecommand \natexlab [1]{#1}%
\providecommand \enquote  [1]{``#1''}%
\providecommand \bibnamefont  [1]{#1}%
\providecommand \bibfnamefont [1]{#1}%
\providecommand \citenamefont [1]{#1}%
\providecommand \href@noop [0]{\@secondoftwo}%
\providecommand \href [0]{\begingroup \@sanitize@url \@href}%
\providecommand \@href[1]{\@@startlink{#1}\@@href}%
\providecommand \@@href[1]{\endgroup#1\@@endlink}%
\providecommand \@sanitize@url [0]{\catcode `\\12\catcode `\$12\catcode
  `\&12\catcode `\#12\catcode `\^12\catcode `\_12\catcode `\%12\relax}%
\providecommand \@@startlink[1]{}%
\providecommand \@@endlink[0]{}%
\providecommand \url  [0]{\begingroup\@sanitize@url \@url }%
\providecommand \@url [1]{\endgroup\@href {#1}{\urlprefix }}%
\providecommand \urlprefix  [0]{URL }%
\providecommand \Eprint [0]{\href }%
\providecommand \doibase [0]{http://dx.doi.org/}%
\providecommand \selectlanguage [0]{\@gobble}%
\providecommand \bibinfo  [0]{\@secondoftwo}%
\providecommand \bibfield  [0]{\@secondoftwo}%
\providecommand \translation [1]{[#1]}%
\providecommand \BibitemOpen [0]{}%
\providecommand \bibitemStop [0]{}%
\providecommand \bibitemNoStop [0]{.\EOS\space}%
\providecommand \EOS [0]{\spacefactor3000\relax}%
\providecommand \BibitemShut  [1]{\csname bibitem#1\endcsname}%
\let\auto@bib@innerbib\@empty
%</preamble>

\bibitem[{\citenamefont{Duerr et~al.}(2016)\citenamefont{Duerr, Kahlhoefer,
  Schmidt-Hoberg, Schwetz, and Vogl}}]{Duerr:2016tmh}
\bibinfo{author}{\bibfnamefont{M.}~\bibnamefont{Duerr}},
  \bibinfo{author}{\bibfnamefont{F.}~\bibnamefont{Kahlhoefer}},
  \bibinfo{author}{\bibfnamefont{K.}~\bibnamefont{Schmidt-Hoberg}},
  \bibinfo{author}{\bibfnamefont{T.}~\bibnamefont{Schwetz}}, \bibnamefont{and}
  \bibinfo{author}{\bibfnamefont{S.}~\bibnamefont{Vogl}},
  \bibinfo{journal}{JHEP} \textbf{\bibinfo{volume}{09}}, \bibinfo{pages}{042}
  (\bibinfo{year}{2016}), \eprint{1606.07609}.

\bibitem[{\citenamefont{Escudero et~al.}(2016)\citenamefont{Escudero, Berlin,
  Hooper, and Lin}}]{Escudero:2016gzx}
\bibinfo{author}{\bibfnamefont{M.}~\bibnamefont{Escudero}},
  \bibinfo{author}{\bibfnamefont{A.}~\bibnamefont{Berlin}},
  \bibinfo{author}{\bibfnamefont{D.}~\bibnamefont{Hooper}}, \bibnamefont{and}
  \bibinfo{author}{\bibfnamefont{M.-X.} \bibnamefont{Lin}},
  \bibinfo{journal}{JCAP} \textbf{\bibinfo{volume}{1612}}, \bibinfo{pages}{029}
  (\bibinfo{year}{2016}), \eprint{1609.09079}.

\bibitem[{\citenamefont{Arcadi et~al.}(2017)\citenamefont{Arcadi, Dutra, Ghosh,
  Lindner, Mambrini, Pierre, Profumo, and Queiroz}}]{Arcadi:2017kky}
\bibinfo{author}{\bibfnamefont{G.}~\bibnamefont{Arcadi}},
  \bibinfo{author}{\bibfnamefont{M.}~\bibnamefont{Dutra}},
  \bibinfo{author}{\bibfnamefont{P.}~\bibnamefont{Ghosh}},
  \bibinfo{author}{\bibfnamefont{M.}~\bibnamefont{Lindner}},
  \bibinfo{author}{\bibfnamefont{Y.}~\bibnamefont{Mambrini}},
  \bibinfo{author}{\bibfnamefont{M.}~\bibnamefont{Pierre}},
  \bibinfo{author}{\bibfnamefont{S.}~\bibnamefont{Profumo}}, \bibnamefont{and}
  \bibinfo{author}{\bibfnamefont{F.~S.} \bibnamefont{Queiroz}}
  (\bibinfo{year}{2017}), \eprint{1703.07364}.

\bibitem[{\citenamefont{Battaglieri et~al.}(2017)}]{Battaglieri:2017aum}
\bibinfo{author}{\bibfnamefont{M.}~\bibnamefont{Battaglieri}}
  \bibnamefont{et~al.} (\bibinfo{year}{2017}), \eprint{1707.04591}.

\bibitem[{\citenamefont{Angloher et~al.}(2016)}]{Angloher:2015ewa}
\bibinfo{author}{\bibfnamefont{G.}~\bibnamefont{Angloher}} \bibnamefont{et~al.}
  (\bibinfo{collaboration}{CRESST}), \bibinfo{journal}{Eur. Phys. J.}
  \textbf{\bibinfo{volume}{C76}}, \bibinfo{pages}{25} (\bibinfo{year}{2016}),
  \eprint{1509.01515}.

\bibitem[{\citenamefont{Angloher et~al.}(2015)}]{Angloher:2015eza}
\bibinfo{author}{\bibfnamefont{G.}~\bibnamefont{Angloher}} \bibnamefont{et~al.}
  (\bibinfo{collaboration}{CRESST}) (\bibinfo{year}{2015}),
  \eprint{1503.08065}.

\bibitem[{\citenamefont{Angloher et~al.}(2017)}]{Angloher:2017sxg}
\bibinfo{author}{\bibfnamefont{G.}~\bibnamefont{Angloher}} \bibnamefont{et~al.}
  (\bibinfo{collaboration}{CRESST}), \bibinfo{journal}{Eur. Phys. J.}
  \textbf{\bibinfo{volume}{C77}}, \bibinfo{pages}{637} (\bibinfo{year}{2017}),
  \eprint{1707.06749}.

\bibitem[{\citenamefont{Petricca et~al.}(2017)}]{Petricca:2017zdp}
\bibinfo{author}{\bibfnamefont{F.}~\bibnamefont{Petricca}} \bibnamefont{et~al.}
  (\bibinfo{collaboration}{CRESST}) (\bibinfo{year}{2017}),
  \eprint{1711.07692}.

\bibitem[{\citenamefont{Aguilar-Arevalo
  et~al.}(2016)}]{Aguilar-Arevalo:2016ndq}
\bibinfo{author}{\bibfnamefont{A.}~\bibnamefont{Aguilar-Arevalo}}
  \bibnamefont{et~al.} (\bibinfo{collaboration}{DAMIC}),
  \bibinfo{journal}{Phys. Rev.} \textbf{\bibinfo{volume}{D94}},
  \bibinfo{pages}{082006} (\bibinfo{year}{2016}), \eprint{1607.07410}.

\bibitem[{\citenamefont{Arnaud et~al.}(2017)}]{Arnaud:2017usi}
\bibinfo{author}{\bibfnamefont{Q.}~\bibnamefont{Arnaud}} \bibnamefont{et~al.}
  (\bibinfo{collaboration}{EDELWEISS}) (\bibinfo{year}{2017}),
  \eprint{1707.04308}.

\bibitem[{\citenamefont{Arnaud et~al.}(2018)}]{Arnaud:2017bjh}
\bibinfo{author}{\bibfnamefont{Q.}~\bibnamefont{Arnaud}} \bibnamefont{et~al.},
  \bibinfo{journal}{Astropart. Phys.} \textbf{\bibinfo{volume}{97}},
  \bibinfo{pages}{54} (\bibinfo{year}{2018}), \eprint{1706.04934}.

\bibitem[{\citenamefont{Agnese et~al.}(2017)}]{Agnese:2016cpb}
\bibinfo{author}{\bibfnamefont{R.}~\bibnamefont{Agnese}} \bibnamefont{et~al.}
  (\bibinfo{collaboration}{SuperCDMS}), \bibinfo{journal}{Phys. Rev.}
  \textbf{\bibinfo{volume}{D95}}, \bibinfo{pages}{082002}
  (\bibinfo{year}{2017}), \eprint{1610.00006}.

\bibitem[{\citenamefont{Essig et~al.}(2012)\citenamefont{Essig, Mardon, and
  Volansky}}]{Essig:2011nj}
\bibinfo{author}{\bibfnamefont{R.}~\bibnamefont{Essig}},
  \bibinfo{author}{\bibfnamefont{J.}~\bibnamefont{Mardon}}, \bibnamefont{and}
  \bibinfo{author}{\bibfnamefont{T.}~\bibnamefont{Volansky}},
  \bibinfo{journal}{Phys. Rev.} \textbf{\bibinfo{volume}{D85}},
  \bibinfo{pages}{076007} (\bibinfo{year}{2012}), \eprint{1108.5383}.

\bibitem[{\citenamefont{Graham et~al.}(2012)\citenamefont{Graham, Kaplan,
  Rajendran, and Walters}}]{Graham:2012su}
\bibinfo{author}{\bibfnamefont{P.~W.} \bibnamefont{Graham}},
  \bibinfo{author}{\bibfnamefont{D.~E.} \bibnamefont{Kaplan}},
  \bibinfo{author}{\bibfnamefont{S.}~\bibnamefont{Rajendran}},
  \bibnamefont{and} \bibinfo{author}{\bibfnamefont{M.~T.}
  \bibnamefont{Walters}}, \bibinfo{journal}{Phys. Dark Univ.}
  \textbf{\bibinfo{volume}{1}}, \bibinfo{pages}{32} (\bibinfo{year}{2012}),
  \eprint{1203.2531}.

\bibitem[{\citenamefont{Essig et~al.}(2016)\citenamefont{Essig,
  Fernandez-Serra, Mardon, Soto, Volansky, and Yu}}]{Essig:2015cda}
\bibinfo{author}{\bibfnamefont{R.}~\bibnamefont{Essig}},
  \bibinfo{author}{\bibfnamefont{M.}~\bibnamefont{Fernandez-Serra}},
  \bibinfo{author}{\bibfnamefont{J.}~\bibnamefont{Mardon}},
  \bibinfo{author}{\bibfnamefont{A.}~\bibnamefont{Soto}},
  \bibinfo{author}{\bibfnamefont{T.}~\bibnamefont{Volansky}}, \bibnamefont{and}
  \bibinfo{author}{\bibfnamefont{T.-T.} \bibnamefont{Yu}},
  \bibinfo{journal}{JHEP} \textbf{\bibinfo{volume}{05}}, \bibinfo{pages}{046}
  (\bibinfo{year}{2016}), \eprint{1509.01598}.

\bibitem[{\citenamefont{Essig et~al.}(2017{\natexlab{a}})\citenamefont{Essig,
  Volansky, and Yu}}]{Essig:2017kqs}
\bibinfo{author}{\bibfnamefont{R.}~\bibnamefont{Essig}},
  \bibinfo{author}{\bibfnamefont{T.}~\bibnamefont{Volansky}}, \bibnamefont{and}
  \bibinfo{author}{\bibfnamefont{T.-T.} \bibnamefont{Yu}},
  \bibinfo{journal}{Phys. Rev.} \textbf{\bibinfo{volume}{D96}},
  \bibinfo{pages}{043017} (\bibinfo{year}{2017}{\natexlab{a}}),
  \eprint{1703.00910}.

\bibitem[{\citenamefont{Guo and McKinsey}(2013)}]{Guo:2013dt}
\bibinfo{author}{\bibfnamefont{W.}~\bibnamefont{Guo}} \bibnamefont{and}
  \bibinfo{author}{\bibfnamefont{D.~N.} \bibnamefont{McKinsey}},
  \bibinfo{journal}{Phys. Rev.} \textbf{\bibinfo{volume}{D87}},
  \bibinfo{pages}{115001} (\bibinfo{year}{2013}), \eprint{1302.0534}.

\bibitem[{\citenamefont{Hochberg
  et~al.}(2016{\natexlab{a}})\citenamefont{Hochberg, Zhao, and
  Zurek}}]{Hochberg:2015pha}
\bibinfo{author}{\bibfnamefont{Y.}~\bibnamefont{Hochberg}},
  \bibinfo{author}{\bibfnamefont{Y.}~\bibnamefont{Zhao}}, \bibnamefont{and}
  \bibinfo{author}{\bibfnamefont{K.~M.} \bibnamefont{Zurek}},
  \bibinfo{journal}{Phys. Rev. Lett.} \textbf{\bibinfo{volume}{116}},
  \bibinfo{pages}{011301} (\bibinfo{year}{2016}{\natexlab{a}}),
  \eprint{1504.07237}.

\bibitem[{\citenamefont{Hochberg
  et~al.}(2016{\natexlab{b}})\citenamefont{Hochberg, Pyle, Zhao, and
  Zurek}}]{Hochberg:2015fth}
\bibinfo{author}{\bibfnamefont{Y.}~\bibnamefont{Hochberg}},
  \bibinfo{author}{\bibfnamefont{M.}~\bibnamefont{Pyle}},
  \bibinfo{author}{\bibfnamefont{Y.}~\bibnamefont{Zhao}}, \bibnamefont{and}
  \bibinfo{author}{\bibfnamefont{K.~M.} \bibnamefont{Zurek}},
  \bibinfo{journal}{JHEP} \textbf{\bibinfo{volume}{08}}, \bibinfo{pages}{057}
  (\bibinfo{year}{2016}{\natexlab{b}}), \eprint{1512.04533}.

\bibitem[{\citenamefont{Hochberg
  et~al.}(2016{\natexlab{c}})\citenamefont{Hochberg, Lin, and
  Zurek}}]{Hochberg:2016ajh}
\bibinfo{author}{\bibfnamefont{Y.}~\bibnamefont{Hochberg}},
  \bibinfo{author}{\bibfnamefont{T.}~\bibnamefont{Lin}}, \bibnamefont{and}
  \bibinfo{author}{\bibfnamefont{K.~M.} \bibnamefont{Zurek}},
  \bibinfo{journal}{Phys. Rev.} \textbf{\bibinfo{volume}{D94}},
  \bibinfo{pages}{015019} (\bibinfo{year}{2016}{\natexlab{c}}),
  \eprint{1604.06800}.

\bibitem[{\citenamefont{Schutz and Zurek}(2016)}]{Schutz:2016tid}
\bibinfo{author}{\bibfnamefont{K.}~\bibnamefont{Schutz}} \bibnamefont{and}
  \bibinfo{author}{\bibfnamefont{K.~M.} \bibnamefont{Zurek}},
  \bibinfo{journal}{Phys. Rev. Lett.} \textbf{\bibinfo{volume}{117}},
  \bibinfo{pages}{121302} (\bibinfo{year}{2016}), \eprint{1604.08206}.

\bibitem[{\citenamefont{Carter et~al.}(2017)\citenamefont{Carter, Hertel,
  Rooks, McClintock, McKinsey, and Prober}}]{Carter:2016wid}
\bibinfo{author}{\bibfnamefont{F.~W.} \bibnamefont{Carter}},
  \bibinfo{author}{\bibfnamefont{S.~A.} \bibnamefont{Hertel}},
  \bibinfo{author}{\bibfnamefont{M.~J.} \bibnamefont{Rooks}},
  \bibinfo{author}{\bibfnamefont{P.~V.~E.} \bibnamefont{McClintock}},
  \bibinfo{author}{\bibfnamefont{D.~N.} \bibnamefont{McKinsey}},
  \bibnamefont{and} \bibinfo{author}{\bibfnamefont{D.~E.}
  \bibnamefont{Prober}}, \bibinfo{journal}{J. Low. Temp. Phys.}
  \textbf{\bibinfo{volume}{186}}, \bibinfo{pages}{183} (\bibinfo{year}{2017}),
  \eprint{1605.00694}.

\bibitem[{\citenamefont{Hochberg
  et~al.}(2017{\natexlab{a}})\citenamefont{Hochberg, Kahn, Lisanti, Tully, and
  Zurek}}]{Hochberg:2016ntt}
\bibinfo{author}{\bibfnamefont{Y.}~\bibnamefont{Hochberg}},
  \bibinfo{author}{\bibfnamefont{Y.}~\bibnamefont{Kahn}},
  \bibinfo{author}{\bibfnamefont{M.}~\bibnamefont{Lisanti}},
  \bibinfo{author}{\bibfnamefont{C.~G.} \bibnamefont{Tully}}, \bibnamefont{and}
  \bibinfo{author}{\bibfnamefont{K.~M.} \bibnamefont{Zurek}},
  \bibinfo{journal}{Phys. Lett.} \textbf{\bibinfo{volume}{B772}},
  \bibinfo{pages}{239} (\bibinfo{year}{2017}{\natexlab{a}}),
  \eprint{1606.08849}.

\bibitem[{\citenamefont{Derenzo et~al.}(2017)\citenamefont{Derenzo, Essig,
  Massari, Soto, and Yu}}]{Derenzo:2016fse}
\bibinfo{author}{\bibfnamefont{S.}~\bibnamefont{Derenzo}},
  \bibinfo{author}{\bibfnamefont{R.}~\bibnamefont{Essig}},
  \bibinfo{author}{\bibfnamefont{A.}~\bibnamefont{Massari}},
  \bibinfo{author}{\bibfnamefont{A.}~\bibnamefont{Soto}}, \bibnamefont{and}
  \bibinfo{author}{\bibfnamefont{T.-T.} \bibnamefont{Yu}},
  \bibinfo{journal}{Phys. Rev.} \textbf{\bibinfo{volume}{D96}},
  \bibinfo{pages}{016026} (\bibinfo{year}{2017}), \eprint{1607.01009}.

\bibitem[{\citenamefont{Hochberg
  et~al.}(2017{\natexlab{b}})\citenamefont{Hochberg, Lin, and
  Zurek}}]{Hochberg:2016sqx}
\bibinfo{author}{\bibfnamefont{Y.}~\bibnamefont{Hochberg}},
  \bibinfo{author}{\bibfnamefont{T.}~\bibnamefont{Lin}}, \bibnamefont{and}
  \bibinfo{author}{\bibfnamefont{K.~M.} \bibnamefont{Zurek}},
  \bibinfo{journal}{Phys. Rev.} \textbf{\bibinfo{volume}{D95}},
  \bibinfo{pages}{023013} (\bibinfo{year}{2017}{\natexlab{b}}),
  \eprint{1608.01994}.

\bibitem[{\citenamefont{Essig et~al.}(2017{\natexlab{b}})\citenamefont{Essig,
  Mardon, Slone, and Volansky}}]{Essig:2016crl}
\bibinfo{author}{\bibfnamefont{R.}~\bibnamefont{Essig}},
  \bibinfo{author}{\bibfnamefont{J.}~\bibnamefont{Mardon}},
  \bibinfo{author}{\bibfnamefont{O.}~\bibnamefont{Slone}}, \bibnamefont{and}
  \bibinfo{author}{\bibfnamefont{T.}~\bibnamefont{Volansky}},
  \bibinfo{journal}{Phys. Rev.} \textbf{\bibinfo{volume}{D95}},
  \bibinfo{pages}{056011} (\bibinfo{year}{2017}{\natexlab{b}}),
  \eprint{1608.02940}.

\bibitem[{\citenamefont{Knapen et~al.}(2017{\natexlab{a}})\citenamefont{Knapen,
  Lin, and Zurek}}]{Knapen:2016cue}
\bibinfo{author}{\bibfnamefont{S.}~\bibnamefont{Knapen}},
  \bibinfo{author}{\bibfnamefont{T.}~\bibnamefont{Lin}}, \bibnamefont{and}
  \bibinfo{author}{\bibfnamefont{K.~M.} \bibnamefont{Zurek}},
  \bibinfo{journal}{Phys. Rev.} \textbf{\bibinfo{volume}{D95}},
  \bibinfo{pages}{056019} (\bibinfo{year}{2017}{\natexlab{a}}),
  \eprint{1611.06228}.

\bibitem[{\citenamefont{Bunting et~al.}(2017)\citenamefont{Bunting, Gratta,
  Melia, and Rajendran}}]{Bunting:2017net}
\bibinfo{author}{\bibfnamefont{P.~C.} \bibnamefont{Bunting}},
  \bibinfo{author}{\bibfnamefont{G.}~\bibnamefont{Gratta}},
  \bibinfo{author}{\bibfnamefont{T.}~\bibnamefont{Melia}}, \bibnamefont{and}
  \bibinfo{author}{\bibfnamefont{S.}~\bibnamefont{Rajendran}},
  \bibinfo{journal}{Phys. Rev.} \textbf{\bibinfo{volume}{D95}},
  \bibinfo{pages}{095001} (\bibinfo{year}{2017}), \eprint{1701.06566}.

\bibitem[{\citenamefont{Budnik et~al.}(2017)\citenamefont{Budnik, Chesnovsky,
  Slone, and Volansky}}]{Budnik:2017sbu}
\bibinfo{author}{\bibfnamefont{R.}~\bibnamefont{Budnik}},
  \bibinfo{author}{\bibfnamefont{O.}~\bibnamefont{Chesnovsky}},
  \bibinfo{author}{\bibfnamefont{O.}~\bibnamefont{Slone}}, \bibnamefont{and}
  \bibinfo{author}{\bibfnamefont{T.}~\bibnamefont{Volansky}}
  (\bibinfo{year}{2017}), \eprint{1705.03016}.

\bibitem[{\citenamefont{Tiffenberg et~al.}(2017)\citenamefont{Tiffenberg,
  Sofo-Haro, Drlica-Wagner, Essig, Guardincerri, Holland, Volansky, and
  Yu}}]{Tiffenberg:2017aac}
\bibinfo{author}{\bibfnamefont{J.}~\bibnamefont{Tiffenberg}},
  \bibinfo{author}{\bibfnamefont{M.}~\bibnamefont{Sofo-Haro}},
  \bibinfo{author}{\bibfnamefont{A.}~\bibnamefont{Drlica-Wagner}},
  \bibinfo{author}{\bibfnamefont{R.}~\bibnamefont{Essig}},
  \bibinfo{author}{\bibfnamefont{Y.}~\bibnamefont{Guardincerri}},
  \bibinfo{author}{\bibfnamefont{S.}~\bibnamefont{Holland}},
  \bibinfo{author}{\bibfnamefont{T.}~\bibnamefont{Volansky}}, \bibnamefont{and}
  \bibinfo{author}{\bibfnamefont{T.-T.} \bibnamefont{Yu}},
  \bibinfo{journal}{Phys. Rev. Lett.} \textbf{\bibinfo{volume}{119}},
  \bibinfo{pages}{131802} (\bibinfo{year}{2017}), \eprint{1706.00028}.

\bibitem[{\citenamefont{Maris et~al.}(2017)\citenamefont{Maris, Seidel, and
  Stein}}]{Maris:2017xvi}
\bibinfo{author}{\bibfnamefont{H.~J.} \bibnamefont{Maris}},
  \bibinfo{author}{\bibfnamefont{G.~M.} \bibnamefont{Seidel}},
  \bibnamefont{and} \bibinfo{author}{\bibfnamefont{D.}~\bibnamefont{Stein}},
  \bibinfo{journal}{Phys. Rev. Lett.} \textbf{\bibinfo{volume}{119}},
  \bibinfo{pages}{181303} (\bibinfo{year}{2017}), \eprint{1706.00117}.

\bibitem[{\citenamefont{Hochberg
  et~al.}(2017{\natexlab{c}})\citenamefont{Hochberg, Kahn, Lisanti, Zurek,
  Grushin, Ilan, Griffin, Liu, and Weber}}]{Hochberg:2017wce}
\bibinfo{author}{\bibfnamefont{Y.}~\bibnamefont{Hochberg}},
  \bibinfo{author}{\bibfnamefont{Y.}~\bibnamefont{Kahn}},
  \bibinfo{author}{\bibfnamefont{M.}~\bibnamefont{Lisanti}},
  \bibinfo{author}{\bibfnamefont{K.~M.} \bibnamefont{Zurek}},
  \bibinfo{author}{\bibfnamefont{A.}~\bibnamefont{Grushin}},
  \bibinfo{author}{\bibfnamefont{R.}~\bibnamefont{Ilan}},
  \bibinfo{author}{\bibfnamefont{S.~M.} \bibnamefont{Griffin}},
  \bibinfo{author}{\bibfnamefont{Z.-F.} \bibnamefont{Liu}}, \bibnamefont{and}
  \bibinfo{author}{\bibfnamefont{S.~F.} \bibnamefont{Weber}}
  (\bibinfo{year}{2017}{\natexlab{c}}), \eprint{1708.08929}.

\bibitem[{\citenamefont{Fichet}(2017)}]{Fichet:2017bng}
\bibinfo{author}{\bibfnamefont{S.}~\bibnamefont{Fichet}}
  (\bibinfo{year}{2017}), \eprint{1705.10331}.

\bibitem[{\citenamefont{{Ruijgrok} et~al.}(1983)\citenamefont{{Ruijgrok},
  {Nijboer}, and {Hoare}}}]{Ruijgrok}
\bibinfo{author}{\bibfnamefont{T.~W.} \bibnamefont{{Ruijgrok}}},
  \bibinfo{author}{\bibfnamefont{B.~R.~A.} \bibnamefont{{Nijboer}}},
  \bibnamefont{and} \bibinfo{author}{\bibfnamefont{M.~R.}
  \bibnamefont{{Hoare}}}, \bibinfo{journal}{Physica A Statistical Mechanics and
  its Applications} \textbf{\bibinfo{volume}{120}}, \bibinfo{pages}{537}
  (\bibinfo{year}{1983}).

\bibitem[{\citenamefont{Vegh}(1983)}]{Vegh}
\bibinfo{author}{\bibfnamefont{L.}~\bibnamefont{Vegh}},
  \bibinfo{journal}{Journal of Physics B: Atomic and Molecular Physics}
  \textbf{\bibinfo{volume}{16}}, \bibinfo{pages}{4175} (\bibinfo{year}{1983}).

\bibitem[{\citenamefont{Baur et~al.}(1983)\citenamefont{Baur, Rosel, and
  Trautmann}}]{Baur}
\bibinfo{author}{\bibfnamefont{G.}~\bibnamefont{Baur}},
  \bibinfo{author}{\bibfnamefont{F.}~\bibnamefont{Rosel}}, \bibnamefont{and}
  \bibinfo{author}{\bibfnamefont{D.}~\bibnamefont{Trautmann}},
  \bibinfo{journal}{Journal of Physics B: Atomic and Molecular Physics}
  \textbf{\bibinfo{volume}{16}}, \bibinfo{pages}{L419} (\bibinfo{year}{1983}).

\bibitem[{\citenamefont{Pindzola et~al.}(2014)\citenamefont{Pindzola, Lee,
  Abdel-Naby, Robicheaux, Colgan, and Ciappina}}]{Pindzola}
\bibinfo{author}{\bibfnamefont{M.~S.} \bibnamefont{Pindzola}},
  \bibinfo{author}{\bibfnamefont{T.~G.} \bibnamefont{Lee}},
  \bibinfo{author}{\bibfnamefont{S.~A.} \bibnamefont{Abdel-Naby}},
  \bibinfo{author}{\bibfnamefont{F.}~\bibnamefont{Robicheaux}},
  \bibinfo{author}{\bibfnamefont{J.}~\bibnamefont{Colgan}}, \bibnamefont{and}
  \bibinfo{author}{\bibfnamefont{M.~F.} \bibnamefont{Ciappina}},
  \bibinfo{journal}{Journal of Physics B: Atomic, Molecular and Optical
  Physics} \textbf{\bibinfo{volume}{47}}, \bibinfo{pages}{195202}
  (\bibinfo{year}{2014}).

\bibitem[{\citenamefont{Sharma}(2017)}]{Sharma:2017fmo}
\bibinfo{author}{\bibfnamefont{P.}~\bibnamefont{Sharma}},
  \bibinfo{journal}{Nucl. Phys.} \textbf{\bibinfo{volume}{A968}},
  \bibinfo{pages}{326} (\bibinfo{year}{2017}).

\bibitem[{\citenamefont{Kouvaris and Pradler}(2017)}]{Kouvaris:2016afs}
\bibinfo{author}{\bibfnamefont{C.}~\bibnamefont{Kouvaris}} \bibnamefont{and}
  \bibinfo{author}{\bibfnamefont{J.}~\bibnamefont{Pradler}},
  \bibinfo{journal}{Phys. Rev. Lett.} \textbf{\bibinfo{volume}{118}},
  \bibinfo{pages}{031803} (\bibinfo{year}{2017}), \eprint{1607.01789}.

\bibitem[{\citenamefont{McCabe}(2017)}]{McCabe:2017rln}
\bibinfo{author}{\bibfnamefont{C.}~\bibnamefont{McCabe}},
  \bibinfo{journal}{Phys. Rev.} \textbf{\bibinfo{volume}{D96}},
  \bibinfo{pages}{043010} (\bibinfo{year}{2017}), \eprint{1702.04730}.

\bibitem[{\citenamefont{Akerib et~al.}(2017{\natexlab{a}})}]{Akerib:2016vxi}
\bibinfo{author}{\bibfnamefont{D.~S.} \bibnamefont{Akerib}}
  \bibnamefont{et~al.} (\bibinfo{collaboration}{LUX}), \bibinfo{journal}{Phys.
  Rev. Lett.} \textbf{\bibinfo{volume}{118}}, \bibinfo{pages}{021303}
  (\bibinfo{year}{2017}{\natexlab{a}}), \eprint{1608.07648}.

\bibitem[{\citenamefont{Aprile et~al.}(2017)}]{Aprile:2017iyp}
\bibinfo{author}{\bibfnamefont{E.}~\bibnamefont{Aprile}} \bibnamefont{et~al.}
  (\bibinfo{collaboration}{XENON}), \bibinfo{journal}{Phys. Rev. Lett.}
  \textbf{\bibinfo{volume}{119}}, \bibinfo{pages}{181301}
  (\bibinfo{year}{2017}), \eprint{1705.06655}.

\bibitem[{\citenamefont{Cui et~al.}(2017)}]{Cui:2017nnn}
\bibinfo{author}{\bibfnamefont{X.}~\bibnamefont{Cui}} \bibnamefont{et~al.}
  (\bibinfo{collaboration}{PandaX-II}), \bibinfo{journal}{Phys. Rev. Lett.}
  \textbf{\bibinfo{volume}{119}}, \bibinfo{pages}{181302}
  (\bibinfo{year}{2017}), \eprint{1708.06917}.

\bibitem[{\citenamefont{Ibe et~al.}(2017)\citenamefont{Ibe, Nakano, Shoji, and
  Suzuki}}]{Ibe:2017yqa}
\bibinfo{author}{\bibfnamefont{M.}~\bibnamefont{Ibe}},
  \bibinfo{author}{\bibfnamefont{W.}~\bibnamefont{Nakano}},
  \bibinfo{author}{\bibfnamefont{Y.}~\bibnamefont{Shoji}}, \bibnamefont{and}
  \bibinfo{author}{\bibfnamefont{K.}~\bibnamefont{Suzuki}}
  (\bibinfo{year}{2017}), \eprint{1707.07258}.

\bibitem[{\citenamefont{Vergados and Ejiri}(2005)}]{Vergados:2004bm}
\bibinfo{author}{\bibfnamefont{J.~D.} \bibnamefont{Vergados}} \bibnamefont{and}
  \bibinfo{author}{\bibfnamefont{H.}~\bibnamefont{Ejiri}},
  \bibinfo{journal}{Phys. Lett.} \textbf{\bibinfo{volume}{B606}},
  \bibinfo{pages}{313} (\bibinfo{year}{2005}), \eprint{hep-ph/0401151}.

\bibitem[{\citenamefont{Moustakidis et~al.}(2005)\citenamefont{Moustakidis,
  Vergados, and Ejiri}}]{Moustakidis:2005gx}
\bibinfo{author}{\bibfnamefont{C.~C.} \bibnamefont{Moustakidis}},
  \bibinfo{author}{\bibfnamefont{J.~D.} \bibnamefont{Vergados}},
  \bibnamefont{and} \bibinfo{author}{\bibfnamefont{H.}~\bibnamefont{Ejiri}},
  \bibinfo{journal}{Nucl. Phys.} \textbf{\bibinfo{volume}{B727}},
  \bibinfo{pages}{406} (\bibinfo{year}{2005}), \eprint{hep-ph/0507123}.

\bibitem[{\citenamefont{Bernabei et~al.}(2007)}]{Bernabei:2007jz}
\bibinfo{author}{\bibfnamefont{R.}~\bibnamefont{Bernabei}}
  \bibnamefont{et~al.}, \bibinfo{journal}{Int. J. Mod. Phys.}
  \textbf{\bibinfo{volume}{A22}}, \bibinfo{pages}{3155} (\bibinfo{year}{2007}),
  \eprint{0706.1421}.

\bibitem[{\citenamefont{Landau and Lifshitz}(1965)}]{Landau:QM}
\bibinfo{author}{\bibfnamefont{L.~D.} \bibnamefont{Landau}} \bibnamefont{and}
  \bibinfo{author}{\bibfnamefont{E.~M.} \bibnamefont{Lifshitz}},
  \emph{\bibinfo{title}{Quantum Mechanics: Non-relativistic theory}}
  (\bibinfo{publisher}{Pergamon Press}, \bibinfo{year}{1965}),
  \bibinfo{note}{{Homework assignment:~ \S~41, Problems 2 \& 3}}.

\bibitem[{\citenamefont{Foot}(2004)}]{Foot:2004pa}
\bibinfo{author}{\bibfnamefont{R.}~\bibnamefont{Foot}}, \bibinfo{journal}{Int.
  J. Mod. Phys.} \textbf{\bibinfo{volume}{D13}}, \bibinfo{pages}{2161}
  (\bibinfo{year}{2004}), \eprint{astro-ph/0407623}.

\bibitem[{\citenamefont{Feng et~al.}(2009)\citenamefont{Feng, Kaplinghat, Tu,
  and Yu}}]{Feng:2009mn}
\bibinfo{author}{\bibfnamefont{J.~L.} \bibnamefont{Feng}},
  \bibinfo{author}{\bibfnamefont{M.}~\bibnamefont{Kaplinghat}},
  \bibinfo{author}{\bibfnamefont{H.}~\bibnamefont{Tu}}, \bibnamefont{and}
  \bibinfo{author}{\bibfnamefont{H.-B.} \bibnamefont{Yu}},
  \bibinfo{journal}{JCAP} \textbf{\bibinfo{volume}{0907}}, \bibinfo{pages}{004}
  (\bibinfo{year}{2009}), \eprint{0905.3039}.

\bibitem[{\citenamefont{McCabe}(2014)}]{McCabe:2013kea}
\bibinfo{author}{\bibfnamefont{C.}~\bibnamefont{McCabe}},
  \bibinfo{journal}{JCAP} \textbf{\bibinfo{volume}{1402}}, \bibinfo{pages}{027}
  (\bibinfo{year}{2014}), \eprint{1312.1355}.

\bibitem[{\citenamefont{{Feist} et~al.}(2008)\citenamefont{{Feist}, {Nagele},
  {Pazourek}, {Persson}, {Schneider}, {Collins}, and {Burgd{\"o}rfer}}}]{Feist}
\bibinfo{author}{\bibfnamefont{J.}~\bibnamefont{{Feist}}},
  \bibinfo{author}{\bibfnamefont{S.}~\bibnamefont{{Nagele}}},
  \bibinfo{author}{\bibfnamefont{R.}~\bibnamefont{{Pazourek}}},
  \bibinfo{author}{\bibfnamefont{E.}~\bibnamefont{{Persson}}},
  \bibinfo{author}{\bibfnamefont{B.~I.} \bibnamefont{{Schneider}}},
  \bibinfo{author}{\bibfnamefont{L.~A.} \bibnamefont{{Collins}}},
  \bibnamefont{and}
  \bibinfo{author}{\bibfnamefont{J.}~\bibnamefont{{Burgd{\"o}rfer}}},
  \bibinfo{journal}{Phys. Rev. A} \textbf{\bibinfo{volume}{77}},
  \bibinfo{eid}{043420} (\bibinfo{year}{2008}), \eprint{0803.0511}.

\bibitem[{\citenamefont{{Liertzer} et~al.}(2012)\citenamefont{{Liertzer},
  {Feist}, {Nagele}, and {Burgd{\"o}rfer}}}]{Liertzer}
\bibinfo{author}{\bibfnamefont{M.}~\bibnamefont{{Liertzer}}},
  \bibinfo{author}{\bibfnamefont{J.}~\bibnamefont{{Feist}}},
  \bibinfo{author}{\bibfnamefont{S.}~\bibnamefont{{Nagele}}}, \bibnamefont{and}
  \bibinfo{author}{\bibfnamefont{J.}~\bibnamefont{{Burgd{\"o}rfer}}},
  \bibinfo{journal}{Physical Review Letters} \textbf{\bibinfo{volume}{109}},
  \bibinfo{eid}{013201} (\bibinfo{year}{2012}), \eprint{1201.5508}.

\bibitem[{\citenamefont{Tucker-Smith and Weiner}(2001)}]{TuckerSmith:2001hy}
\bibinfo{author}{\bibfnamefont{D.}~\bibnamefont{Tucker-Smith}}
  \bibnamefont{and} \bibinfo{author}{\bibfnamefont{N.}~\bibnamefont{Weiner}},
  \bibinfo{journal}{Phys.Rev.} \textbf{\bibinfo{volume}{D64}},
  \bibinfo{pages}{043502} (\bibinfo{year}{2001}), \eprint{hep-ph/0101138}.

\bibitem[{\citenamefont{Chepel and Araujo}(2013)}]{Chepel:2012sj}
\bibinfo{author}{\bibfnamefont{V.}~\bibnamefont{Chepel}} \bibnamefont{and}
  \bibinfo{author}{\bibfnamefont{H.}~\bibnamefont{Araujo}},
  \bibinfo{journal}{JINST} \textbf{\bibinfo{volume}{8}},
  \bibinfo{pages}{R04001} (\bibinfo{year}{2013}), \eprint{1207.2292}.

\bibitem[{\citenamefont{Sorensen}(2015)}]{Sorensen:2014sla}
\bibinfo{author}{\bibfnamefont{P.}~\bibnamefont{Sorensen}},
  \bibinfo{journal}{Phys. Rev.} \textbf{\bibinfo{volume}{D91}},
  \bibinfo{pages}{083509} (\bibinfo{year}{2015}), \eprint{1412.3028}.

\bibitem[{\citenamefont{McCabe}(2016)}]{McCabe:2015eia}
\bibinfo{author}{\bibfnamefont{C.}~\bibnamefont{McCabe}},
  \bibinfo{journal}{JCAP} \textbf{\bibinfo{volume}{1605}}, \bibinfo{pages}{033}
  (\bibinfo{year}{2016}), \eprint{1512.00460}.

\bibitem[{\citenamefont{Szydagis et~al.}(2011)\citenamefont{Szydagis, Barry,
  Kazkaz, Mock, Stolp, Sweany, Tripathi, Uvarov, Walsh, and
  Woods}}]{Szydagis:2011tk}
\bibinfo{author}{\bibfnamefont{M.}~\bibnamefont{Szydagis}},
  \bibinfo{author}{\bibfnamefont{N.}~\bibnamefont{Barry}},
  \bibinfo{author}{\bibfnamefont{K.}~\bibnamefont{Kazkaz}},
  \bibinfo{author}{\bibfnamefont{J.}~\bibnamefont{Mock}},
  \bibinfo{author}{\bibfnamefont{D.}~\bibnamefont{Stolp}},
  \bibinfo{author}{\bibfnamefont{M.}~\bibnamefont{Sweany}},
  \bibinfo{author}{\bibfnamefont{M.}~\bibnamefont{Tripathi}},
  \bibinfo{author}{\bibfnamefont{S.}~\bibnamefont{Uvarov}},
  \bibinfo{author}{\bibfnamefont{N.}~\bibnamefont{Walsh}}, \bibnamefont{and}
  \bibinfo{author}{\bibfnamefont{M.}~\bibnamefont{Woods}},
  \bibinfo{journal}{JINST} \textbf{\bibinfo{volume}{6}},
  \bibinfo{pages}{P10002} (\bibinfo{year}{2011}), \eprint{1106.1613}.

\bibitem[{\citenamefont{Szydagis et~al.}(2013)\citenamefont{Szydagis, Fyhrie,
  Thorngren, and Tripathi}}]{Szydagis:2013sih}
\bibinfo{author}{\bibfnamefont{M.}~\bibnamefont{Szydagis}},
  \bibinfo{author}{\bibfnamefont{A.}~\bibnamefont{Fyhrie}},
  \bibinfo{author}{\bibfnamefont{D.}~\bibnamefont{Thorngren}},
  \bibnamefont{and} \bibinfo{author}{\bibfnamefont{M.}~\bibnamefont{Tripathi}},
  \bibinfo{journal}{JINST} \textbf{\bibinfo{volume}{8}},
  \bibinfo{pages}{C10003} (\bibinfo{year}{2013}), \eprint{1307.6601}.

\bibitem[{\citenamefont{Lenardo et~al.}(2015)\citenamefont{Lenardo, Kazkaz,
  Manalaysay, Mock, Szydagis, and Tripathi}}]{Lenardo:2014cva}
\bibinfo{author}{\bibfnamefont{B.}~\bibnamefont{Lenardo}},
  \bibinfo{author}{\bibfnamefont{K.}~\bibnamefont{Kazkaz}},
  \bibinfo{author}{\bibfnamefont{A.}~\bibnamefont{Manalaysay}},
  \bibinfo{author}{\bibfnamefont{J.}~\bibnamefont{Mock}},
  \bibinfo{author}{\bibfnamefont{M.}~\bibnamefont{Szydagis}}, \bibnamefont{and}
  \bibinfo{author}{\bibfnamefont{M.}~\bibnamefont{Tripathi}},
  \bibinfo{journal}{IEEE Trans. Nucl. Sci.} \textbf{\bibinfo{volume}{62}},
  \bibinfo{pages}{3387} (\bibinfo{year}{2015}), \eprint{1412.4417}.

\bibitem[{\citenamefont{Akerib et~al.}(2016{\natexlab{a}})}]{Akerib:2015wdi}
\bibinfo{author}{\bibfnamefont{D.~S.} \bibnamefont{Akerib}}
  \bibnamefont{et~al.} (\bibinfo{collaboration}{LUX}), \bibinfo{journal}{Phys.
  Rev.} \textbf{\bibinfo{volume}{D93}}, \bibinfo{pages}{072009}
  (\bibinfo{year}{2016}{\natexlab{a}}), \eprint{1512.03133}.

\bibitem[{\citenamefont{Goetzke et~al.}(2017)\citenamefont{Goetzke, Aprile,
  Anthony, Plante, and Weber}}]{Goetzke:2016lfg}
\bibinfo{author}{\bibfnamefont{L.~W.} \bibnamefont{Goetzke}},
  \bibinfo{author}{\bibfnamefont{E.}~\bibnamefont{Aprile}},
  \bibinfo{author}{\bibfnamefont{M.}~\bibnamefont{Anthony}},
  \bibinfo{author}{\bibfnamefont{G.}~\bibnamefont{Plante}}, \bibnamefont{and}
  \bibinfo{author}{\bibfnamefont{M.}~\bibnamefont{Weber}},
  \bibinfo{journal}{Phys. Rev.} \textbf{\bibinfo{volume}{D96}},
  \bibinfo{pages}{103007} (\bibinfo{year}{2017}), \eprint{1611.10322}.

\bibitem[{\citenamefont{Boulton et~al.}(2017)}]{Boulton:2017hub}
\bibinfo{author}{\bibfnamefont{E.~M.} \bibnamefont{Boulton}}
  \bibnamefont{et~al.}, \bibinfo{journal}{JINST} \textbf{\bibinfo{volume}{12}},
  \bibinfo{pages}{P08004} (\bibinfo{year}{2017}), \eprint{1705.08958}.

\bibitem[{\citenamefont{Akerib et~al.}(2017{\natexlab{b}})}]{Akerib:2017hph}
\bibinfo{author}{\bibfnamefont{D.~S.} \bibnamefont{Akerib}}
  \bibnamefont{et~al.} (\bibinfo{collaboration}{LUX})
  (\bibinfo{year}{2017}{\natexlab{b}}), \eprint{1709.00800}.

\bibitem[{\citenamefont{Aprile et~al.}(2018{\natexlab{a}})}]{Aprile:2017xxh}
\bibinfo{author}{\bibfnamefont{E.}~\bibnamefont{Aprile}} \bibnamefont{et~al.}
  (\bibinfo{collaboration}{XENON}), \bibinfo{journal}{Phys. Rev.}
  \textbf{\bibinfo{volume}{D97}}, \bibinfo{pages}{092007}
  (\bibinfo{year}{2018}{\natexlab{a}}), \eprint{1709.10149}.

\bibitem[{\citenamefont{Akerib et~al.}(2016{\natexlab{b}})}]{Akerib:2015rjg}
\bibinfo{author}{\bibfnamefont{D.~S.} \bibnamefont{Akerib}}
  \bibnamefont{et~al.} (\bibinfo{collaboration}{LUX}), \bibinfo{journal}{Phys.
  Rev. Lett.} \textbf{\bibinfo{volume}{116}}, \bibinfo{pages}{161301}
  (\bibinfo{year}{2016}{\natexlab{b}}), \eprint{1512.03506}.

\bibitem[{\citenamefont{Aprile et~al.}(2018{\natexlab{b}})}]{Aprile:2018dbl}
\bibinfo{author}{\bibfnamefont{E.}~\bibnamefont{Aprile}} \bibnamefont{et~al.}
  (\bibinfo{collaboration}{XENON}) (\bibinfo{year}{2018}{\natexlab{b}}),
  \eprint{1805.12562}.

\bibitem[{\citenamefont{Barlow}(1990)}]{Barlow:1990vc}
\bibinfo{author}{\bibfnamefont{R.~J.} \bibnamefont{Barlow}},
  \bibinfo{journal}{Nucl. Instrum. Meth.} \textbf{\bibinfo{volume}{A297}},
  \bibinfo{pages}{496} (\bibinfo{year}{1990}).

\bibitem[{\citenamefont{Cowan et~al.}(2011)\citenamefont{Cowan, Cranmer, Gross,
  and Vitells}}]{Cowan:2010js}
\bibinfo{author}{\bibfnamefont{G.}~\bibnamefont{Cowan}},
  \bibinfo{author}{\bibfnamefont{K.}~\bibnamefont{Cranmer}},
  \bibinfo{author}{\bibfnamefont{E.}~\bibnamefont{Gross}}, \bibnamefont{and}
  \bibinfo{author}{\bibfnamefont{O.}~\bibnamefont{Vitells}},
  \bibinfo{journal}{Eur. Phys. J.} \textbf{\bibinfo{volume}{C71}},
  \bibinfo{pages}{1554} (\bibinfo{year}{2011}), \bibinfo{note}{[Erratum: Eur.
  Phys. J.C73,2501(2013)]}, \eprint{1007.1727}.

\bibitem[{\citenamefont{Mount et~al.}(2017)}]{Mount:2017qzi}
\bibinfo{author}{\bibfnamefont{B.~J.} \bibnamefont{Mount}} \bibnamefont{et~al.}
  (\bibinfo{year}{2017}), \eprint{1703.09144}.

\bibitem[{\citenamefont{Aprile et~al.}(2016{\natexlab{a}})}]{Aprile:2015uzo}
\bibinfo{author}{\bibfnamefont{E.}~\bibnamefont{Aprile}} \bibnamefont{et~al.}
  (\bibinfo{collaboration}{XENON}), \bibinfo{journal}{JCAP}
  \textbf{\bibinfo{volume}{1604}}, \bibinfo{pages}{027}
  (\bibinfo{year}{2016}{\natexlab{a}}), \eprint{1512.07501}.

\bibitem[{\citenamefont{Akerib et~al.}(2018)}]{Akerib:2018lyp}
\bibinfo{author}{\bibfnamefont{D.~S.} \bibnamefont{Akerib}}
  \bibnamefont{et~al.} (\bibinfo{collaboration}{LUX-ZEPLIN})
  (\bibinfo{year}{2018}), \eprint{1802.06039}.

\bibitem[{\citenamefont{Kaplinghat et~al.}(2014)\citenamefont{Kaplinghat,
  Tulin, and Yu}}]{Kaplinghat:2013yxa}
\bibinfo{author}{\bibfnamefont{M.}~\bibnamefont{Kaplinghat}},
  \bibinfo{author}{\bibfnamefont{S.}~\bibnamefont{Tulin}}, \bibnamefont{and}
  \bibinfo{author}{\bibfnamefont{H.-B.} \bibnamefont{Yu}},
  \bibinfo{journal}{Phys. Rev.} \textbf{\bibinfo{volume}{D89}},
  \bibinfo{pages}{035009} (\bibinfo{year}{2014}), \eprint{1310.7945}.

\bibitem[{\citenamefont{Kahlhoefer
  et~al.}(2017{\natexlab{a}})\citenamefont{Kahlhoefer, Schmidt-Hoberg, and
  Wild}}]{Kahlhoefer:2017umn}
\bibinfo{author}{\bibfnamefont{F.}~\bibnamefont{Kahlhoefer}},
  \bibinfo{author}{\bibfnamefont{K.}~\bibnamefont{Schmidt-Hoberg}},
  \bibnamefont{and} \bibinfo{author}{\bibfnamefont{S.}~\bibnamefont{Wild}},
  \bibinfo{journal}{JCAP} \textbf{\bibinfo{volume}{1708}}, \bibinfo{pages}{003}
  (\bibinfo{year}{2017}{\natexlab{a}}), \eprint{1704.02149}.

\bibitem[{\citenamefont{Kahlhoefer
  et~al.}(2017{\natexlab{b}})\citenamefont{Kahlhoefer, Kulkarni, and
  Wild}}]{Kahlhoefer:2017ddj}
\bibinfo{author}{\bibfnamefont{F.}~\bibnamefont{Kahlhoefer}},
  \bibinfo{author}{\bibfnamefont{S.}~\bibnamefont{Kulkarni}}, \bibnamefont{and}
  \bibinfo{author}{\bibfnamefont{S.}~\bibnamefont{Wild}}
  (\bibinfo{year}{2017}{\natexlab{b}}), \eprint{1707.08571}.

\bibitem[{\citenamefont{An et~al.}(2015)\citenamefont{An, Pospelov, Pradler,
  and Ritz}}]{An:2014twa}
\bibinfo{author}{\bibfnamefont{H.}~\bibnamefont{An}},
  \bibinfo{author}{\bibfnamefont{M.}~\bibnamefont{Pospelov}},
  \bibinfo{author}{\bibfnamefont{J.}~\bibnamefont{Pradler}}, \bibnamefont{and}
  \bibinfo{author}{\bibfnamefont{A.}~\bibnamefont{Ritz}},
  \bibinfo{journal}{Phys. Lett.} \textbf{\bibinfo{volume}{B747}},
  \bibinfo{pages}{331} (\bibinfo{year}{2015}), \eprint{1412.8378}.

\bibitem[{\citenamefont{Essig et~al.}(2013)\citenamefont{Essig, Mardon,
  Papucci, Volansky, and Zhong}}]{Essig:2013vha}
\bibinfo{author}{\bibfnamefont{R.}~\bibnamefont{Essig}},
  \bibinfo{author}{\bibfnamefont{J.}~\bibnamefont{Mardon}},
  \bibinfo{author}{\bibfnamefont{M.}~\bibnamefont{Papucci}},
  \bibinfo{author}{\bibfnamefont{T.}~\bibnamefont{Volansky}}, \bibnamefont{and}
  \bibinfo{author}{\bibfnamefont{Y.-M.} \bibnamefont{Zhong}},
  \bibinfo{journal}{JHEP} \textbf{\bibinfo{volume}{11}}, \bibinfo{pages}{167}
  (\bibinfo{year}{2013}), \eprint{1309.5084}.

\bibitem[{\citenamefont{Lees et~al.}(2017)}]{Lees:2017lec}
\bibinfo{author}{\bibfnamefont{J.~P.} \bibnamefont{Lees}} \bibnamefont{et~al.}
  (\bibinfo{collaboration}{BaBar}), \bibinfo{journal}{Phys. Rev. Lett.}
  \textbf{\bibinfo{volume}{119}}, \bibinfo{pages}{131804}
  (\bibinfo{year}{2017}), \eprint{1702.03327}.

\bibitem[{\citenamefont{Andreas et~al.}(2012)\citenamefont{Andreas, Niebuhr,
  and Ringwald}}]{Andreas:2012mt}
\bibinfo{author}{\bibfnamefont{S.}~\bibnamefont{Andreas}},
  \bibinfo{author}{\bibfnamefont{C.}~\bibnamefont{Niebuhr}}, \bibnamefont{and}
  \bibinfo{author}{\bibfnamefont{A.}~\bibnamefont{Ringwald}},
  \bibinfo{journal}{Phys. Rev.} \textbf{\bibinfo{volume}{D86}},
  \bibinfo{pages}{095019} (\bibinfo{year}{2012}), \eprint{1209.6083}.

\bibitem[{\citenamefont{Izaguirre et~al.}(2013)\citenamefont{Izaguirre,
  Krnjaic, Schuster, and Toro}}]{Izaguirre:2013uxa}
\bibinfo{author}{\bibfnamefont{E.}~\bibnamefont{Izaguirre}},
  \bibinfo{author}{\bibfnamefont{G.}~\bibnamefont{Krnjaic}},
  \bibinfo{author}{\bibfnamefont{P.}~\bibnamefont{Schuster}}, \bibnamefont{and}
  \bibinfo{author}{\bibfnamefont{N.}~\bibnamefont{Toro}},
  \bibinfo{journal}{Phys. Rev.} \textbf{\bibinfo{volume}{D88}},
  \bibinfo{pages}{114015} (\bibinfo{year}{2013}), \eprint{1307.6554}.

\bibitem[{\citenamefont{Batell et~al.}(2014)\citenamefont{Batell, Essig, and
  Surujon}}]{Batell:2014mga}
\bibinfo{author}{\bibfnamefont{B.}~\bibnamefont{Batell}},
  \bibinfo{author}{\bibfnamefont{R.}~\bibnamefont{Essig}}, \bibnamefont{and}
  \bibinfo{author}{\bibfnamefont{Z.}~\bibnamefont{Surujon}},
  \bibinfo{journal}{Phys. Rev. Lett.} \textbf{\bibinfo{volume}{113}},
  \bibinfo{pages}{171802} (\bibinfo{year}{2014}), \eprint{1406.2698}.

\bibitem[{\citenamefont{Aprile et~al.}(2016{\natexlab{b}})}]{Aprile:2016wwo}
\bibinfo{author}{\bibfnamefont{E.}~\bibnamefont{Aprile}} \bibnamefont{et~al.}
  (\bibinfo{collaboration}{XENON}), \bibinfo{journal}{Phys. Rev.}
  \textbf{\bibinfo{volume}{D94}}, \bibinfo{pages}{092001}
  (\bibinfo{year}{2016}{\natexlab{b}}), \bibinfo{note}{[Erratum: Phys.
  Rev.D95,no.5,059901(2017)]}, \eprint{1605.06262}.

\bibitem[{\citenamefont{Boehm et~al.}(2013)\citenamefont{Boehm, Dolan, and
  McCabe}}]{Boehm:2013jpa}
\bibinfo{author}{\bibfnamefont{C.}~\bibnamefont{Boehm}},
  \bibinfo{author}{\bibfnamefont{M.~J.} \bibnamefont{Dolan}}, \bibnamefont{and}
  \bibinfo{author}{\bibfnamefont{C.}~\bibnamefont{McCabe}},
  \bibinfo{journal}{JCAP} \textbf{\bibinfo{volume}{1308}}, \bibinfo{pages}{041}
  (\bibinfo{year}{2013}), \eprint{1303.6270}.

\bibitem[{\citenamefont{Knapen et~al.}(2017{\natexlab{b}})\citenamefont{Knapen,
  Lin, and Zurek}}]{Knapen:2017xzo}
\bibinfo{author}{\bibfnamefont{S.}~\bibnamefont{Knapen}},
  \bibinfo{author}{\bibfnamefont{T.}~\bibnamefont{Lin}}, \bibnamefont{and}
  \bibinfo{author}{\bibfnamefont{K.~M.} \bibnamefont{Zurek}}
  (\bibinfo{year}{2017}{\natexlab{b}}), \eprint{1709.07882}.

\bibitem[{\citenamefont{Ade et~al.}(2016)}]{Ade:2015xua}
\bibinfo{author}{\bibfnamefont{P.~A.~R.} \bibnamefont{Ade}}
  \bibnamefont{et~al.} (\bibinfo{collaboration}{Planck}),
  \bibinfo{journal}{Astron. Astrophys.} \textbf{\bibinfo{volume}{594}},
  \bibinfo{pages}{A13} (\bibinfo{year}{2016}), \eprint{1502.01589}.

\bibitem[{\citenamefont{Akimov et~al.}(2017)}]{Akimov:2017ade}
\bibinfo{author}{\bibfnamefont{D.}~\bibnamefont{Akimov}} \bibnamefont{et~al.}
  (\bibinfo{collaboration}{COHERENT}), \bibinfo{journal}{Science}
  \textbf{\bibinfo{volume}{357}}, \bibinfo{pages}{1123} (\bibinfo{year}{2017}),
  \eprint{1708.01294}.

\bibitem[{\citenamefont{Verbus et~al.}(2017)}]{Verbus:2016sgw}
\bibinfo{author}{\bibfnamefont{J.~R.} \bibnamefont{Verbus}}
  \bibnamefont{et~al.}, \bibinfo{journal}{Nucl. Instrum. Meth.}
  \textbf{\bibinfo{volume}{A851}}, \bibinfo{pages}{68} (\bibinfo{year}{2017}),
  \eprint{1608.05309}.

\bibitem[{\citenamefont{Barbeau et~al.}(2007)\citenamefont{Barbeau, Collar, and
  Whaley}}]{Barbeau:2007qh}
\bibinfo{author}{\bibfnamefont{P.~S.} \bibnamefont{Barbeau}},
  \bibinfo{author}{\bibfnamefont{J.~I.} \bibnamefont{Collar}},
  \bibnamefont{and} \bibinfo{author}{\bibfnamefont{P.~M.}
  \bibnamefont{Whaley}}, \bibinfo{journal}{Nucl. Instrum. Meth.}
  \textbf{\bibinfo{volume}{A574}}, \bibinfo{pages}{385} (\bibinfo{year}{2007}),
  \eprint{nucl-ex/0701011}.
  
  \bibitem[{fie()}]{fields}
\bibinfo{note}{J.~Balajthy, APS April Meeting 2018,
  \url{https://meetings.aps.org/Meeting/APR18/Session/J09.3}}.

\bibitem[{\citenamefont{Akerib et~al.}(2018{\natexlab{a}})}]{Akerib:2017vbi}
\bibinfo{author}{\bibfnamefont{D.~S.} \bibnamefont{Akerib}}
  \bibnamefont{et~al.} (\bibinfo{collaboration}{LUX}), \bibinfo{journal}{Phys.
  Rev.} \textbf{\bibinfo{volume}{D97}}, \bibinfo{pages}{102008}
  (\bibinfo{year}{2018}{\natexlab{a}}), \eprint{1712.05696}.

\bibitem[{\citenamefont{{Aalbers}}(2018)}]{aablers}
\bibinfo{author}{\bibfnamefont{J.}~\bibnamefont{{Aalbers}}}, Ph.D. thesis,
  \bibinfo{school}{University of Amsterdam} (\bibinfo{year}{2018}).


\end{thebibliography}
\end{document}